\documentstyle[prl,aps,epsf,12pt]{revtex}
\begin{document}
\draft

\title{Non-Abelian Holonomy of BCS and SDW Quasiparticles}
\author{Eugene Demler and Shou-Cheng Zhang}
\address{
Department of Physics,
Stanford University
Stanford, CA 94305
}
\date{\today}
\maketitle
\begin{abstract}
In this work we investigate properties of fermions in the 
$SO(5)$ theory of high $T_c$ superconductivity. We show that the
adiabatic time evolution of a $SO(5)$ superspin vector 
leads to a non-Abelian $SU(2)$ holonomy of the $SO(5)$ spinor
states. Physically, this non-trivial holonomy arises from the
non-zero overlap between the SDW and BCS quasi-particle states.
While the usual Berry's phase of a $SO(3)$ spinor is described by 
a Dirac magnetic monopole at the degeneracy point, the non-Abelian
holonomy of a $SO(5)$ spinor is described by a Yang monopole at the
degeneracy point, and is deeply related to the existence of the
second Hopf map from $S^7$ to $S^4$. We conclude this work by
extending the bosonic $SO(5)$ nonlinear $\sigma$ model to include
the fermionic states around the gap nodes as 4 component Dirac
fermions coupled to $SU(2)$ gauge fields in 2+1 dimensions. 

\end{abstract}

\newpage

\section{Introduction}
\label{sec:I}

Recently, a unified theory based on $SO(5)$ symmetry between
antiferromagnetism (AF) and $d$ wave superconductivity (dSC)
has been proposed\cite{science} for
the high $T_c$ cuprates. Initially, this theory was formulated in terms
of a nonlinear $\sigma$ model which describes the effective bosonic
degrees of freedom below the pseudogap temperature. 
This theory gives a unified description of the
high $T_c$ phase diagram and offers a natural explanation of the
$\pi$ resonance mode\cite{prl95,prb98} observed in the $YBCO$ 
superconductors. With the exception of exact microscopic $SO(5)$
models \cite{prl98-2,chenley,Burgess,szh}, both numerical 
investigations \cite{prl97-2,Eder} and the experimental 
proposals \cite{prl98-2,Burgess,vortex,junction2} have primarily 
focused on the bosonic sector of the $SO(5)$ theory.

However, it is clear that a complete theory of high $T_c$ superconductivity
has to properly account for the fermionic degrees of freedom as well.
Some key experiments on the pseudogap physics, {\it e.g.} the ARPES 
experiments, primarily probe the single electron properties
rather than the collective modes. Within the $SO(5)$ theory, the 
pseudogap regime is identified with the fluctuations of the orientation
of the $SO(5)$ superspin vector. Therefore, it is essential to understand
how the fluctuations of the superspin couple to single particle fermionic
degrees of freedom. Because of the $d$ wave nodes, there are 
fermionic excitations with low energy, and they make important
contributions to thermodynamics and to the damping of the collective modes.

Motivated by these considerations, we investigate the fermionic
sector of the $SO(5)$ theory. Our main interest is to understand
how the fermionic degrees of freedom are coupled to the bosonic
$SO(5)$ superspin vector, and investigate the novel topological
properties of this coupling. Rather surprisingly, we find that this
coupling leads to a $SU(2)$ non-Abelian holonomy of the fermionic 
states. The fermionic states in the $SO(5)$ theory are nothing
but the familiar SDW (spin-density-wave) and BCS (Bardeen-Cooper-Schrieffer)
quasiparticles relevant for the AF insulator phase and the dSC phase. 
Since AF and dSC states have very different physical properties, one
would naively expect the SDW and BCS quasiparticle states to be 
orthogonal to each other. The central result of our work shows that
this is not the case. {\it The SDW and BCS quasi-particle states have
nonzero overlaps, and this overlap
defines a precise connection for their adiabatic evolution.}
This simple physical property of non-orthogonality of the SDW and BCS 
quasi-particles leads to an extremely rich mathematical structure.
In particular, 
because of the inherent degeneracy of the SDW and BCS quasi-particle
states, the adiabatic evolution is non-Abelian, and can interchange
the degenerate states upon a cyclic evolution returning to the origin.

Our current investigation is also motivated by the question ``what is
special about the $SO(5)$ symmetry group?". Is it simply introduced
as a convenient mathematical description of the AF and dSC phases
in a unified framework, or is there something deeper which explains
its uniqueness and calls for its natural emergence? In this work,
we give partial mathematical answers to these probing questions,
as we embark on 
a journey through some of the most elegant and 
beautiful mathematical concepts in group theory, differential geometry,
topology and division algebra, 
which, as shall see, are unified by the $SO(5)$ group
in a profound and unique way. The following table summarizes
some of the main mathematical properties of the holonomy of a $SO(5)$
spinor, in comparison with the Berry's phase of a $SO(3)$ spinor:     

\vspace{.2in}
\begin{tabular}{|c|c|c|}    \hline
\hspace{2cm}         &$SO(3)$ Spinor &$SO(5)$ Spinor\\ \hline
Holonomy &$U(1)$ Berry's Phase   &$SU(2)$ Wilczek-Zee Holonomy\\  \hline
Singularity at Degeneracy Point &Dirac Monopole   &Yang Monopole\\  \hline
Topological Invariant &First Chern Number   &Second Chern Number\\  \hline
Wigner-von Neumann Class &Unitary   &Symplectic\\  \hline
Associated Hopf Maps &$S^3$ to $S^2$   &$S^7$ to $S^4$\\  \hline
Associated Division Algebra & Complex Numbers   & Hamilton Numbers (Quaternions)\\  \hline
\end{tabular} 
\vspace{.2in}

Soon after Berry's discovery of the adiabatic phase
\cite{Berry,Simon}, Wilczek and  Zee\cite{WilczekZee}
generalized this concept to the non-Abelian holonomy in a quantum
system with degeneracy. In a sense, the 
most natural generalization of the concept of the 
Berry's phase of a $SO(3)$ spinor is the $SU(2)$ non-Abelian holonomy
of a $SO(5)$ spinor. The usual Abelian Berry's phase
has its mathematical origin in the first Hopf map from a 
three sphere $S^3$ to a two sphere $S^2$. Similarly,
the non-trivial holonomy of a $SO(5)$ spinor is
deeply related to the existence of the second Hopf map from the
seven sphere $S^7$ to the four sphere $S^4$, the later being the order
parameter space of the $SO(5)$ theory. The generalization from the
first Hopf map $S^3 \rightarrow S^2$ to the second Hopf map 
$S^7 \rightarrow S^4$ is uniquely related to the generalization
from complex numbers to Hamilton numbers (quaternions). 
In mathematics, there exist only three kinds of division algebra,
namely complex numbers, Hamilton numbers (quaternions) and 
Cayley numbers (octernions). These three classes of division
algebra lead to three types of Hopf maps\cite{steenrod}, 
$S^3\rightarrow S^2$, $S^7\rightarrow S^4$
and $S^{11}\rightarrow S^7$. Octernions and the third Hopf map
are not useful in physics because of the lack of associativity.
This makes the second Hopf map and its associated
non-Abelian holonomy of a $SO(5)$ spinor an 
essentially unique generalization of the usual Abelian Berry's
phase. 

The novel topological property of a $SO(5)$ spinor enables us to
extend the bosonic $SO(5)$ non-linear $\sigma$ model to include
the fermionic states. Because of the $d$ wave nodes, the low energy
electronic states can be best described as Dirac fermions in $2+1$
dimension. The $SU(2)$ non-Abelian
holonomy uniquely determines the coupling of these Dirac fermions
to the fluctuation of the $SO(5)$ superspin order parameter, and
takes the form of a minimal coupling to a $SU(2)$ gauge field. With the
inclusion of low energy fermionic modes, the formulation of the $SO(5)$ theory
is essentially complete. The present formalism can be used to 
systematically investigate the fermionic properties which result
from the interplay between AF and dSC, for example, the pseudogap
physics, fermionic quantum numbers and bound state inside the
$SO(5)$ vortices and  junctions, electronic states in the stripe
phase etc. The non-Abelian holonomy of the BCS and SDW quasi-particles
discussed in this work could also be observed directly in experiments. 
However, in the paper, we shall only restrict ourselves
to the mathematical formulation of the theory, physical
application of the present formalism will be discussed in future works.

Some of the mathematical properties presented in this paper have been
discussed previously in other contexts. General background on Berry's
phase and its various generalizations have been collected in an
authorative reprint volume edited
by Shapere and Wilczek\cite{ShapereWilczek}. 
Minami\cite{Minami} studied the connection between the second
Hopf map and the Yang monopole. Wu and Zee\cite{WuZee} studied a 
$SO(5)$ nonlinear $\sigma$ model in seven dimensions and discussed 
the connection between the second Hopf map and quaternionic 
algebra. Avron {\it et al} \cite{Avron1,Avron2} discussed the general
settings of 
Berry's phase in a fermionic time reversal invariant systems
and the connections to Wigner von Neumann classes.
An interesting example of the the non-Abelian Berry's phase
has been studied by Mathur, who showed that such term appears in the
adiabatic effective Hamiltonian for the orbital motion of a Dirac
electron and leads to spin-orbit interaction and Thomas
precession \cite{Mathur}. Shankar 
and Mathur later identified the non-Abelian Berry vector potential in
this problem with that of a meron \cite{Shankar}.
While some of the mathematical concepts discussed in this paper may
not sound familiar to some readers, we shall present our paper in an 
essentially self-contained fashion which should be accessible without
much background mathematical knowledge.

\section{Holonomy of a SO(5) Spinor}

\subsection{Adiabatic Evolution of a $SO(5)$ Spinor}
\label{AdEvol}

Abelian Berry's phase or $U(1)$ holonomy is a familiar object in
physics. A typical example comes from considering a spin
$\frac{1}{2}$ particle coupled to a magnetic field\cite{Berry,Simon,Stone}. 
The Hamiltonian for this problem is a simple $2\times 2$ matrix, given by
\begin{eqnarray}
{\cal H}_{\alpha\beta} = \Delta~ n_a(t) \sigma^a_{\alpha \beta} 
\label{spinhalf}
\end{eqnarray}
where $n_a(t)$ ($a=x,y,z$) is a three dimensional unit vector, 
$\sigma^a$ are the three $2\times 2$ Pauli matrices and $\Delta$
is the Zeeman energy gap. The projection operator 
\begin{eqnarray}
P_{\pm}
= \frac{1}{2}( 1 \pm n_a \sigma^a ) 
\end{eqnarray}
projects onto the subspace of {\it non-degenerate} eigenvalues
$\pm \Delta$. Under the assumptions of adiabatic evolution,
we can choose the instantaneous eigenstate corresponding to 
the eigenvalue $+\Delta$ as 
\begin{eqnarray}
|\Psi^{(1)}_+(t)\rangle = \frac{1}{N_1} P_{+}(n(t)) 
\left( \begin{array}{l}1\\0 \end{array} \right)
\end{eqnarray}
where $N_1$ is a normalization factor, which ensures $\langle
\Psi^{(1)}_+(t) | \Psi^{(1)}_+(t)\rangle =1 $.
The Berry's phase is defined in terms of the non-vanishing overlap
between these instantaneous eigenstates:
\begin{eqnarray}
\gamma^{(1)}_+(t) = -i \int_0^t d\tau 
\langle\Psi^{(1)}_+(\tau)|\frac{d}{d\tau}|\Psi^{(1)}_+(\tau)\rangle
\end{eqnarray}
This phase factor can in turn be expressed as a line integral over
the vector potential of a Dirac monopole:
\begin{eqnarray}
\gamma^{(1)}_+(t) = \int_0^t d\tau  A_+^a(n(\tau)) dn_a(\tau)
\end{eqnarray}
where the vector potential
\begin{eqnarray}
A_+^a dn_a = \frac{1}{2(1+n_z)} (n_x dn_y - n_y dn_x)
\label{A+}
\end{eqnarray}
is singular at the south pole $n_z=-1$. 

Similarly, one can choose another class of instantaneous eigenstates
corresponding to the $+\Delta$ eigenvalue, defined as 
\begin{eqnarray}
|\Psi^{(2)}_+(t)\rangle = \frac{1}{N_2} P_{+}(n(t)) 
\left( \begin{array}{l}0\\1 \end{array} \right)
\end{eqnarray}
The Berry's phase associated with this class of instantaneous eigenstates
is given by 
\begin{eqnarray}
\gamma^{(2)}_+(t) = -i \int_0^t d\tau 
\langle\Psi^{(2)}_+(\tau)|\frac{d}{d\tau}|\Psi^{(2)}_+(\tau)\rangle
= \int_0^t d\tau {\tilde A}_+^a(n(\tau)) dn_a(\tau)
\end{eqnarray}
where the vector potential
\begin{eqnarray}
{\tilde A}_+^a dn_a = -\frac{1}{2(1-n_z)} (n_x dn_y - n_y dn_x)
\label{tildeA+}
\end{eqnarray}
is singular at the north pole $\theta=0$. 
$A_+(n)$ and ${\tilde A}_+(n)$ define the two nonsingular ``patches" 
of the monopole section in the sense of Wu and Yang\cite{WuYang}, their 
difference is a pure gauge in the overlapping equatorial region $S^1$,
where $n_z=0$ and $n_x^2 + n_y^2=1$: 
\begin{eqnarray}
d W = A_+^a dn_a - {\tilde A}_+^a dn_a = n_x dn_y - n_y dn_x  
\label{U1gte} 
\end{eqnarray}
 
Obviously, we can also define the Berry's phase of the instantaneous
eigenstates with eigenvalue $-\Delta$, and find their associated
gauge potentials $A_-(n)$ and ${\tilde A}_-(n)$. They correspond to
a Dirac monopole with opposite magnetic charge.

How does the concept of a Berry's phase of a $SO(3)$ spinor 
generalize to the case of a $SO(5)$ spinor? The Hamiltonian for
a $SO(5)$ spinor coupled to a $SO(5)$ unit vector $n_a$ ($ a=1,..,5$),
called superspin in the $SO(5)$ theory, is given by a simple 
generalization of (\ref{spinhalf}):
\begin{eqnarray}
{\cal H}_{\alpha\beta} = \Delta~ n_a(t) \Gamma^a_{\alpha \beta} 
\label{spinorH}
\end{eqnarray}
where the five $4\times 4$ Dirac $\Gamma$ matrices are given by
\begin{eqnarray}
 \Gamma^1\!=\! \left( \begin{array}{cc}
               0          & -i\sigma_y  \\
               i\sigma_y  & 0            \end{array} \right)
 \Gamma^{(2,3,4)} \!=\! \left( \begin{array}{cc}
             \vec \sigma  & 0  \\
                  0       &  ^t\vec \sigma  \end{array} \right)
 \Gamma^5 \!=\! \left( \begin{array}{cc}
                0         & \sigma_y  \\
                \sigma_y  & 0            \end{array} \right)
\end{eqnarray}
Here $\vec \sigma=(\sigma_x,\sigma_y,\sigma_z)$ are the usual
Pauli matrices and $^t\vec \sigma$ denotes their transposition.
$\Delta$ has the physical interpretation of a SDW gap energy
when the superspin vector $n_a$ points in the $2,3,4$ direction,
and a BCS gap energy when it points in the $1,5$ direction.
 
Here we see a crucial difference between the $SO(3)$ spinor Hamiltonian
(\ref{spinhalf}) and the $SO(5)$ spinor Hamiltonian (\ref{spinorH}).
The instantaneous eigenvalues of both Hamiltonians are $\pm \Delta$.
However, the eigenvalues are non-degenerate for the $SO(3)$ case, but doubly
degenerate for the $SO(5)$ case. 
For example,
when the $n$-field points along $n_4$, Hamiltonian
(\ref{spinorH}) is diagonal, with 
${\cal H}_{11}= {\cal H}_{33}=- {\cal H}_{22}= -{\cal H}_{44}= \Delta$. 
It is then immediately obvious that states in the subspace
of  $|\Psi_{2} \rangle$ and $|\Psi_{4} \rangle$ have energy
$-\Delta$, and states in the subspace
of  $|\Psi_{1} \rangle$ and $|\Psi_{3} \rangle$ have energy
$+\Delta$. 

Wilczek and Zee \cite{WilczekZee} 
pointed out that adiabatic Hamiltonians with degeneracies can 
have non-Abelian
holonomies and may be characterized by the non-Abelian gauge connections. 
Let us imagine a slow (adiabatic) rotation of
the $SO(5)$ unit vector field when it traces a closed cycle on $S^4$. 
What happens to our our spinor states?
In the adiabatic approximation we can assume that when we start with a
spinor state in the low energy subspace it will never be excited into
the upper subspace. However the low energy subspace is
two-dimensional. In the course of adiabatic rotation, the spinor
state is always some linear combination of the two degenerate low
energy states. Different instantaneous eigenstates are related to each
other by a $2\times 2$ unitary matrix. Moreover, 
after the five-vector returns to its original direction the spinor
state does not necessarily return into itself  but
into some linear combination of the initial state with its
degenerate orthogonal state. Therefore spinors in $SO(5)$ have a
non-Abelian $SU(2)$ holonomy which has its origin in the double
degeneracy of spinor states and leads to many novel
phenomena that we discuss below.    

Following the discussions of the Abelian Berry's phase, we introduce
the projection operators
\begin{eqnarray}
P_{\pm} = \frac{1}{2}(1 \pm n_a \Gamma^a)\ \ ;\ \ 
P_\pm^2 = P_\pm\ \ ;\ \  
{\cal H} P_\pm = \pm \Delta P_\pm
\label{projector4}
\end{eqnarray}
It is convenient to parameterize the four sphere $S^4$ as
\begin{eqnarray}
n_1 &=& sin \theta_1 sin \theta_2 sin \phi_2 \nonumber\\  
n_2 &=& sin \theta_1 cos \theta_2 cos \phi_1 \nonumber\\  
n_3 &=& sin \theta_1 cos \theta_2 sin \phi_1 \nonumber\\  
n_4 &=& cos \theta_1 \nonumber\\  
n_5 &=& sin \theta_1 sin \theta_2 cos \phi_2 
\label{parametrize}
\end{eqnarray}
With these representations, we can choose one class of instantaneous
two dimensional basis states corresponding to the $+\Delta$ eigenvalue as 
\begin{eqnarray}
| \Psi_+^{(1)}\rangle &=& 
\frac{1}{N_1} P_+ {\bf e}_1 \nonumber\\
&=& \frac{1}{\sqrt{2(1+cos \theta_1)}}
\left( 1+ cos \theta_1, sin \theta_1 cos \theta_2 e^{i \phi_1},
 0, i sin \theta_1 sin \theta_2 e^{i \phi_2} \right)^{T}
\nonumber\\
| \Psi_+^{(3)}\rangle &=& \frac{1}{N_3}
P_+ {\bf e}_3 \nonumber\\
&=& \frac{1}{\sqrt{2(1 + cos \theta_1)}}
\left( 0, i sin \theta_1 sin \theta_2  e^{- i \phi_2},
 1+ cos \theta_1, sin \theta_1 sin \theta_2 e^{-i \phi_1}
\right)^{T}
\end{eqnarray}
where we defined unit vectors ${\bf e}_i$ such that $({\bf e}_i)_j =
\delta_{ij} $ (for example ${\bf e}_1 = (1,0,0,0)^{T}$)
and normalization factors $N_1$ and $N_3$ are chosen such that
$\langle \Psi_+^{(i)} | \Psi_+^{(j)} \rangle = \delta_{ij}$. 
These states obey
$ {\cal H}(n) | \Psi_+^{(i)} \rangle= +\Delta~ |\Psi_+^{(i)} \rangle $ by
construction. From this basis of instantaneous eigenstates we obtain
the following non-Abelian holonomy matrix:
\begin{eqnarray}
i {\cal A}_+^a d n_a = \left[ \begin{array}{cc}
\langle \Psi_+^{(1)}| d \Psi_+^{(1)} \rangle & 
\langle \Psi_+^{(1)}| d \Psi_+^{(3)} \rangle \\
\langle \Psi_+^{(3)}| d \Psi_+^{(1)} \rangle & 
\langle \Psi_+^{(3)} | d \Psi_+^{(3)} \rangle
\end{array} \right]
\label{A-definition}
\end{eqnarray}

Alternatively, we could have used the following set of instantaneous
eigenstate basis for the $+\Delta$ eigenvalue,
\begin{eqnarray}
| \Psi_+^{(2)}\rangle = \frac{1}{N_2}~
P_+ {\bf e}_2
\ \ ; \ \
| \Psi_+^{(4)}\rangle = \frac{1}{N_4}~
P_+ {\bf e}_4
\end{eqnarray}
From this alternative basis set we obtain the following non-Abelian
holonomy matrix:
\begin{eqnarray}
i \tilde{\cal A}_+^a d n_a = \left[ \begin{array}{cc}
\langle \Psi_+^{(2)}| d \Psi_+^{(2)} \rangle & 
\langle \Psi_+^{(2)}| d \Psi_+^{(4)} \rangle \\
\langle \Psi_+^{(4)}| d \Psi_+^{(2)} \rangle & 
\langle \Psi_+^{(4)} | d \Psi_+^{(4)} \rangle
\end{array} \right]
\label{DifA-definition}
\end{eqnarray}

\subsection{Adiabatic Connection and Yang's Monopole}
\label{SU2}

From the discussions in the previous section, we see that the non-Abelian
vector potential ${\cal A}^a$ and $\tilde{\cal A}^a$ are direct 
generalizations of the vector potentials $A^a$ and ${\tilde A}^a$
in the Abelian Berry's phase problem, which are the vector potentials of  
a Dirac magnetic monopole. Furthermore, the $2\times 2$ matrix 
$ {\cal A}^a d n_a$ is traceless and hermitian, thus defining a
$SU(2)$ gauge potential. Therefore, it is natural to ask
if ${\cal A}^a$ and $\tilde{\cal A}^a$ are the $SU(2)$ generalizations
of the Abelian Dirac monopole potential.

Direct computation of ${\cal A}^a$ gives
\begin{eqnarray}
i [{\cal A}_+^a]_{11} d n_a &=& \frac{i}{2( 1+ sin \theta cos \alpha)} \left(  sin^2
\theta_1 cos^2 \theta_2~ d \phi_1 +  sin^2 \theta_1 sin^2 \theta_2~ d\phi_2 \right) 
\nonumber\\&=&
\frac{i}{2( 1+ n_4)} \left( n_2 dn_3 - n_3 d n_2 - n_1 dn_5 + n_5 d n_1 \right)
\nonumber\\
i [{\cal A}_+^a]_{12} d n_a &=& \frac{i e^{ -i ( \phi_1 + \phi_2)}}{2(
  1+ cos \theta_1 )} \left(  sin^2 \theta_1~ d \theta_2 +  i sin^2 \theta_1
sin \theta_2 cos \theta_2~ d \phi_1 - i sin^2 \theta_1
sin \theta_2 cos \theta_2~ d \phi_2 \right) 
\nonumber\\&=& \frac{1}{2( 1+ n_4)} ( n_2 d n_5 - n_5 d n_2 + n_1 d n_3
- n_3 d n_1 
\nonumber\\&&\hspace{2cm} + i~ ( n_1 d n_2 - n_2 d n_1 + n_5 d n_3 -
n_3 d n_5 )) 
\nonumber\\
i [{\cal A}_+^a]_{21} d n_a &=& i [{\cal A}_+^a]^{*}_{12} d n_a 
\nonumber\\
i [{\cal A}_+^a]_{22} d n_a &=& - i [{\cal A}_+^a]_{11} d n_a
\label{calA+} 
\end{eqnarray}
The explicit form of $\tilde{\cal A}_+$ can also be obtained and it
may be shown that it has singularity at $n_4 = 1$. Therefore, while
${\cal A}_+$ 
is non-singular except at the  
``south pole" $n_4=-1$, $\tilde{\cal A}_+$ is 
non-singular except at the ``north pole" $n_4=1$. In the overlapping
region, they are related to each other by a pure non-Abelian gauge
transformation: 
\begin{eqnarray}
 \tilde{{\cal A}}_+^a  d n_a &=& {\cal W}^{-1} (  {\cal A}_+^a d n_a )~
{\cal W} -i~
{\cal W}^{-1} d {\cal W} \nonumber\\  
{\cal W} &=& \frac{1}{\sqrt{1-n_4^2}} \left[ \begin{array}{cc} n_2 - i
    n_3 & n_1 - i n_5 \\ -n_1 -i n_5 & n_2 + i n_3 \end{array} \right]
\end{eqnarray}
Similarly, we can define the vector potentials
${\cal A}_-$ and $\tilde{\cal A}_-$ associated with the $-\Delta$
eigenvalue. These gauge fields turn out to be exactly the 
vector potential of a Yang monopole!

In 1978, Yang found a beautiful generalization of the concept of
the Dirac magnetic monopole \cite{Yang,Yang2}. While the Dirac monopole is a
$SO(3)$  
symmetric point singularity in the three dimensional space, the 
Yang monopole is a $SO(5)$ symmetric point singularity in the 
five dimensional space. The 
Dirac monopole defines a topologically nontrivial $U(1)$ fiber
bundle over the two sphere $S^2$, the Yang monopole defines a 
topologically nontrivial $SU(2)$ fiber bundle over the four sphere
$S^4$. The best way to understand a magnetic monopole is through the
concept of Wu-Yang section. For the case of a Dirac monopole, one
divides the $S^2$ sphere into a ``northern hemisphere" and a 
``southern hemisphere", over which two different non-singular
vector potentials $A^a$ and ${\tilde A}^a$ are defined. 
The ``overlapping region" between the two hemispheres is the
equator, or $S^1$. In this overlapping region, the two $U(1)$
gauge potentials can be ``patched together" non-trivially $\tilde{A}^a
= A^a - i~ W^{-1} \partial^a W$. Since 
the mapping from the overlapping region $S^1$ to the group space
$U(1)$ can in general be characterized by a winding number, 
this integer uniquely defines the non-trivial $U(1)$ fiber bundle 
over $S^2$. This integer is called the first Chern number, and is defined
by 
\begin{eqnarray}
2 \pi c_1 &=& \oint_{S^2} F^{ab} d\sigma_{ab} \nonumber\\
&=& \oint_{S^1} A^a dn_a - 
\oint_{S^1} \tilde{A}^a dn_a  \nonumber\\
&=& i~ \oint_{S^1} W^{-1} \partial^a W dn_a
\label{Chern1}
\end{eqnarray}
where $F^{ab}=\partial^a A^b - \partial^b A^a
=\partial^a {\tilde A}^b - \partial^b {\tilde A}^a$ is the field
strength, $W$ is a gauge transformation between $A$ and $\tilde{A}$
  defined in equation (\ref{U1gte}) and $ d\sigma_{ab}$ is a surface
  element of a 2-sphere. Here and everywhere else in this paper we
  normalize surface element on a $n$-sphere $S^n$ by 
  requiring that when integrated over a unit sphere it gives
  the correct surface area
$\oint_{S_n} \epsilon^{a_1 \dots a_{n+1}} \hat{x}_{a_1} d \sigma_{a_2
  \dots a_{n+1}} = \Omega_n = 2
  \pi^{(n+1)/2}/\Gamma(\frac{n+1}{2}) $.
 Readers familiar with differential geometry will easily recognize $d
  \sigma_{a_2 
  \dots a_{n+1}} = d x_{a_2} \wedge \dots \wedge d x_{a_{n+1}} $. For
  the Berry's connection of a spin one half particle,  
we obtain from formula (\ref{Chern1}) $c_1=\pm 1$ for the $A_+^a$ and
  $A_-^a$ vector potentials respectively.

From the point of view of Wu-Yang section, it is straightforward 
to see why a $SU(2)$ point monopole has to be enclosed by $S^4$.
If we view the ``north pole" as $n_4=+1$ and the ``south pole" as
$n_4=-1$, then the ``overlapping region" between the two hemispheres
is given by $n_4=0$, and is a three sphere $S^3$ defined by
$n_1^2+n_2^2+n_3^2+n_5^2=1$. Since the mapping from $S^3$ to the group
manifold of $SU(2)$ can be characterized by a winding number, 
one can define a topologically non-trivial
``patching" between the ``northern hemisphere" $SU(2)$ gauge potential
$\cal A$ and the ``southern hemisphere" $SU(2)$ gauge potential
$\tilde{\cal A}$. This integer is called the second Chern number and
is defined by the generalization of (\ref{Chern1}):
\begin{eqnarray}
8 \pi^2 c_2 &=& Tr \oint_{S^4} {\cal F}^{ab} {\cal F}^{cd} d\sigma_{abcd}
\nonumber\\ 
& = & 2~ Tr \oint_{S^3} ( {\cal A}^{a} \partial^b {\cal A}^c   -
\frac{2i}{3}~ {\cal A}^a 
{\cal A}^b {\cal A}^c) d\sigma_{abc}  
-  ({\cal A}\rightarrow {\tilde{\cal A}}) \nonumber\\
 & = & \frac{8}{3}~ Tr \oint_{S^3} ( {\cal W}^{-1} \partial^a {\cal W} {\cal
   W}^{-1} \partial^b {\cal W} {\cal W}^{-1} \partial^c {\cal W} )
 d\sigma_{abc} 
\label{Chern2}
\end{eqnarray}
where ${\cal F}^{ab}=\partial^a {\cal A}^b - \partial^b {\cal A}^c + i
[{\cal A}^a, {\cal A}^b]$ is the non-Abelian field strength 
associate with the vector potential ${\cal A}^a$ or $\tilde {\cal
  A}^a$ \cite{Shankar,Jackiw}. 

From the above equation, we easily recognize the integral over the
three sphere as the non-Abelian Chern-Simons term. One may wonder
in what sense is the Chern-Simons term a natural generalization
of the Bohm-Aharonov type of line integral defined in (\ref{Chern1}).
To make the connection between (\ref{Chern1}) and (\ref{Chern2}) more
precise, one can introduce an {\it Abelian} totally antisymmetric
three-index gauge field defined by
\begin{eqnarray}
A^{abc} &=& 2~ Tr ({\cal A}^{[a} \partial^b {\cal A}^{c]} -\frac{2i}{3}~
{\cal A}^{[a} {\cal A}^b {\cal A}^{c]}) \nonumber\\
&=& \frac{1}{3} \sum_{P(abc)}
(-)^{P}~ Tr
({\cal A}^{P(a)}\partial^{P(b)} {\cal A}^{P(c)} -\frac{2i}{3}~
{\cal A}^{P(a)} {\cal A}^{P(b)} {\cal A}^{P(c)} )
\label{threeindex}
\end{eqnarray}
and its associated four-index field strength $F^{abcd}$. In this
representation, the second Chern number appears to be a direct
extension of the first Chern number:
\begin{eqnarray}
8 \pi^2 c_2 = \oint_{S^3} A^{abc} d\sigma_{abc} 
-   \oint_{S^3} {\tilde A}^{abc} d\sigma_{abc}
= \oint_{S^4} F^{abcd} d\sigma_{abcd}
\end{eqnarray}

Since the $SU(2)$ holonomy connection we found explicitly 
in (\ref{calA+}) is $SO(5)$ symmetric and gives $c_2 =\pm 1$
for the ${\cal A}_+$ and ${\cal A}_-$ gauge potentials respectively, 
it can be uniquely identified with the gauge potential of a Yang
monopole. The introduction of the Yang monopole into the $SO(5)$
theory is a important step in describing the fermionic degrees of
freedom. A single bosonic $SO(5)$ rotor has only the fully symmetric
traceless tensor representations of the $SO(5)$ group. However, a 
single rotor with a Yang monopole at the center contain 
{\it all} irreducible representations of the $SO(5)$ group, including
the fermionic spinor sector \cite{Yang2}.

\subsection{Spinor Rotation Matrix}
\label{SpinorRot}
Having established the topological structure of the non-Abelian 
holonomy, let us now go back to our problem of the coupled $SO(5)$ 
spinor and vector and try to actually solve the time-dependent 
Hamiltonian
\begin{eqnarray}
{\cal H}(t) = \Delta~ n_a~(t) \Psi^{\dagger}_{\alpha}
\Gamma^a_{\alpha\beta} \Psi_{\beta} 
\end{eqnarray}
For the purpose of later application to BCS and SDW quasi-particles,
we use here the second-quantized notations, so $\Psi_{\alpha}$'s are now
operators that may be thought of as creating spinor states out of
the vacuum. Time dependence of these operators may be found from
the Heisenberg equation of motion
by taking a commutator of $\Psi_{\alpha}$ with the Hamiltonian and we shall 
assume that $\Psi_{\alpha}$'s anticommute with each other. 
We decompose $\Psi_{\alpha}$ using 
\begin{eqnarray} 
\Psi_{\alpha}(t) = S_{\alpha \beta}(t) \Phi_{\beta}(t)
\label{decomposition} 
\end{eqnarray}
where matrix $S_{\alpha \beta}(t)$ is a unitary matrix and $\Phi_{\beta}$
is another spinor.
The Heisenberg equation of motion for $\Psi_{\alpha}$ can be expressed as
\begin{eqnarray} 
\dot{S}_{\alpha \beta}\Phi_{\beta} + S_{\alpha \beta}\dot{\Phi}_{\beta} =
- i~\Delta~ n_a \Gamma^a_{\alpha\beta} S_{\beta\gamma}\Phi_{\gamma} 
\label{heisenberg} 
\end{eqnarray}
If we choose $\Phi_{\beta}$ such that its time evolution is given by the time
independent Hamiltonian
\begin{eqnarray}
{\cal H}_0 = \Delta~ \Phi^{\dagger}_{\alpha}
\Gamma^4_{\alpha\beta} \Phi_{\beta} 
\label{phiH}
\end{eqnarray} 
then (\ref{heisenberg}) can be solved if $S_{\alpha \beta}$
satisfies the following two conditions
\begin{eqnarray}
n_a(t) &S^{\dagger}(t)& \Gamma^a S(t) = \Gamma^4 
\label{S-condition1}\\
&S^{\dagger}(t)&  \partial_t S(t) = 0
\label{S-condition2}
\end{eqnarray}
The meaning of decomposition (\ref{decomposition}) is clear. At each
moment $t$ we use matrix $S_{\alpha\beta}$ to rotate the spinor from its
instantaneous direction $n_a(t)$ to a fixed direction
$n_4$. However this does not define the matrix $S_{\alpha\beta}$
uniquely. There are 
two states pointing along $n_4$ and two orthogonal states pointing
in the opposite direction. So the first equation 
(\ref{S-condition1}) only fixes $S_{\alpha\beta}$ up to arbitrary
rotation within 
these two pairs. The second equation (\ref{S-condition2}) combined
with the first one determines $S_{\alpha\beta}(t)$ 
uniquely. 

Let us now see how we can construct the spinor rotation matrix that
satisfies both conditions (\ref{S-condition1}),
(\ref{S-condition2}). For a moment we 
forget about the second condition and try to find {\it any } 
unitary matrix $\tilde{S}_{\alpha\beta}$ such that
(\ref{S-condition1}) is satisfied.
Matrix $\Gamma^4$
has two eigenvectors with eigenvalues +1 and two eigenvectors with
eigenvalues -1. Therefore if we find two orthogonal eigenvectors of
$n_a \Gamma^a$ with eigenvalues +1 and two with eigenvalues -1,
they will specify a necessary rotation for us ( eigenvectors that
correspond to different eigenvalues are orthogonal ). These
eigenvectors may be easily found using projection operators 
$P_{\pm}$ defined in (\ref{projector4}). 

We take 
$
P_{+} {\bf e}_1
$
,
$
P_{+} {\bf e}_3
$
,
$
P_{-} {\bf e}_2
$
, 
$
P_{-} {\bf e}_4
$
and call the normalized vectors $y^{(i)}$. They satisfy 
\begin{eqnarray}
y^{(i)*}_{\alpha} y^{(j)}_{\alpha} &=& \delta_{ij} \nonumber\\
 ( n_a \Gamma^a)_{\alpha\beta} y^{(i)}_{\beta} &=& \lambda_i
 y^{(i)}_{\alpha} 
\end{eqnarray}
where $\lambda_i=+1$ when $i=1,3$ and $\lambda_i=-1$ when $i=2,4$. 
It is then clear that we should define $\tilde{S}_{\alpha \beta} =
y^{(\beta)}_{\alpha}$.
For example, if we parameterize superspin as in (\ref{parametrize}),
the matrix $\tilde{S}_{\alpha \beta}$ is explicitly given by 
\begin{eqnarray}
\tilde{S}_{\alpha \beta} = \left[ \begin{array}{cccc}
cos\frac{\theta_1}{2} & - sin\frac{\theta_1}{2} cos \theta_2 e^{-i \phi_1}&
 0 &  i sin\frac{\theta_1}{2} sin\theta_2 e^{-i \phi_2} \\
sin\frac{\theta_1}{2} cos \theta_2 e^{i \phi_1}& cos\frac{\theta_1}{2}&
i sin\frac{\theta_1}{2} sin\theta_2 e^{-i \phi_2} & 0 \\
0 & i sin\frac{\theta_1}{2} sin\theta_2 e^{i \phi_2} &
cos\frac{\theta_1}{2} & - sin\frac{\theta_1}{2} cos \theta_2 e^{i
  \phi_1} \\
i sin\frac{\theta_1}{2} sin\theta_2 e^{i \phi_2} & 0 &
sin\frac{\theta_1}{2} cos \theta_2 e^{-i \phi_1} &
cos\frac{\theta_1}{2} \end{array}
\right]
\label{tildeS}
\end{eqnarray}

Now we recall about the second condition (\ref{S-condition2}) on
$S_{\alpha\beta}$.  Here it is important to keep in mind that we are working within
adiabatic approximation where transitions between states of different
energies are forbidden. Therefore in equation (\ref{S-condition2}) we
don't need to consider the full 4$\times$4 equation but only two
2$\times$2 equations (one of the positive energy subspace with
components 1 and 3 and the other of the negative energy subspace with
components 2 and 4). For notational convenience let us introduce a
matrix 
\begin{eqnarray}
G = 
\left( \begin{array}{cccc}
1 & 0 &  0 &  0 \\
0 & 0 &  1 &  0 \\
0 & 1 &  0 &  0 \\
0 & 0 &  0 &  1  \end{array}
\right)\ \ ; \ \ G^2 = 1
\end{eqnarray}
From Section \ref{BerrysPhase} we know that
within positive and negative energy subspaces $\tilde{S}^{\dagger} d
\tilde{S} $ 
corresponds to an infinitesimal $SU(2)$ rotation, therefore, it can be
expressed as 
\begin{eqnarray}
\tilde{S}^{\dagger} d \tilde{S} = G 
\left( \begin{array}{cc}
i d {\cal A}_+ & 0 \\
0 & i d {\cal A}_-  \end{array}
\right) G
\label{GStilda}
\end{eqnarray}
where the $2\times 2$ $SU(2)$ gauge matrices $d {\cal A}_+$ and $d
{\cal A}_-$ 
are specified in section \ref{BerrysPhase}. 
Using $\tilde S$, ${\cal A}_+$, and ${\cal A}_-$  we can
finally express $S$ as  
\begin{eqnarray}
S(t) = \tilde{S}(t) G 
\left( \begin{array}{cc}
{\cal U}_+ & 0 \\
0 & {\cal U}_-  \end{array}
\right) G
\label{Scomb}
\end{eqnarray}
where
\begin{eqnarray}
d {\cal U}_{\pm}(t) = - i~ {\cal U}_{\pm}(t) d {\cal A}_{\pm}\ \ ; \ \
{\cal U}_{\pm}(t) = T e^{ - i \int_0^t {\cal A}_{\pm}^a(\tau) d n_a(\tau)}
\label{Urot} 
\end{eqnarray}
$T$ is the time ordering symbol and
matrices ${\cal U}_{\pm}$ are the finite $SU(2)$
holonomy matrices for the upper energy and lower energy subspaces. 
It is easy to check that the resulting
$S(t)$ satisfies both equations
(\ref{S-condition1}) and 
(\ref{S-condition2}). 

\subsection{Hamilton's Number, Hopf's Map} 
\label{HopfMap}

In 1843, while walking along the Bridge of Brougham, 
Hamilton discovered a beautiful generalization from complex
numbers to quaternions; 
In 1931, the same year when Dirac formulated the theory of the
magnetic monopoles, Hopf introduced the concept of Hopf maps.
These two great discoveries in algebra and geometry are intimately
related to each other\cite{steenrod,Minami,WuZee}, and they provide a deep
explanation for the non-Abelian holonomy of a $SO(5)$ spinor.

The first Hopf map is a mapping from
$S^3$ to $S^2$ and is related to Dirac's magnetic monopole,
while the second Hopf map is a mapping from $S^7$ to $S^4$, and is deeply 
related to Yang's monopole. $S^3$ is locally isomorphic to 
$S^2\times U(1)$, where we can view $S^2$ as the sphere enclosing the
Dirac monopole and $U(1)$ is the gauge field due to a Dirac monopole.
In the presence of a Dirac monopole, the $U(1)$ bundle over $S^2$ is
topologically non-trivial. However since $S^3$ is parallisible, one
can use the first Hopf map to define a nonsingular vector potential
due to Dirac's monopole everywhere on $S^3$. Similarly, $S^7$ is 
locally isomorphic to $S^4\times SU(2)$, and one can use the second 
Hopf map to define a non-singular $SU(2)$ gauge field everywhere on
$S^7$.

It is simple to define the first Hopf map in physicist's language.
One introduces a two component complex scalar field with a constraint,
\begin{eqnarray}
z_\alpha = \left( \begin{array}{c} z_1 \\ z_2 \end{array} \right) 
\hspace{2cm} z^{\dagger} z = | z_1 |^2 + | z_2 |^2 = 1
\label{znorm}
\end{eqnarray} 
Then the first Hopf map is defined as
\begin{eqnarray}
\eta_1:~ n_a = z^{\dagger} \sigma_a z 
\label{eta1}
\end{eqnarray} 
where $\sigma$'s are the usual Pauli matrices.
Condition (\ref{znorm}) ensures that
$n^2 =1$, so we have a mapping from $z$ with three independent
components ( $S^3$ ) into $n^a$ with two independent components ( $S^2$
) $\eta_1:~ S^3 \rightarrow S^2$.

If we start from the singular gauge potential $A_+^a$ 
on $S^2$ defined by (\ref{A+})
and substitute for $n_a$ the definition of the Hopf map (\ref{eta1}), 
we find an induced gauge potential on $S^3$
\begin{eqnarray}
\omega = i A^a dn_a = \omega^\dagger_\alpha dz_\alpha + \omega_\alpha
dz^\dagger_\alpha   
= -\frac{1}{2} [z_2^\dagger dz_2 - z_2 dz^\dagger_2 - \frac{|z_2|^2}{|z_1|^2}
(z_1^\dagger dz_1 - z_1 dz^\dagger_1)]
\end{eqnarray} 
This $U(1)$ gauge potential is singular on a great circle $|z_1|=0$ of
$S^3$. However, unlike its counterpart $A_+^a$ on $S^2$ , the singularity of
$\omega$ can be completely removed by a gauge transformation:
\begin{eqnarray}
u = \frac{z_1}{|z_1|} \ \ ; \ \
\Omega = \omega +  u^{-1} du
= - \frac{1}{2} (z_\alpha^\dagger dz_\alpha - z_\alpha
dz^\dagger_\alpha)
\label{Omega-eq}  
\end{eqnarray} 
We see that the new $U(1)$ gauge potential $\Omega$ is non-singular
everywhere on $S^3$. Similarly, the gauge potential $\tilde\omega$
induced by ${\tilde A}_+^a$ defined in (\ref{tildeA+})
is singular on the great circle 
$|z_2|=0$ of $S^3$. This singularity can also be removed by a 
gauge transformation using ${\tilde u} = \frac{z_2}{|z_2|}$, giving the
same non-singular $U(1)$ gauge potential $\Omega$.

\begin{figure*}[h]
\centerline{\epsfysize=3.8cm 
\epsfbox{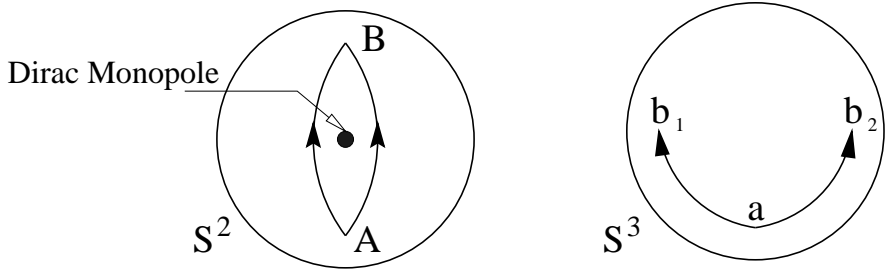}
}
\caption{Transport of a vector on $S^2$ defines transport of a
  spinor on $S^3$. The final spinor state, i.e. the final point on
  $S^3$ depends not only the final point of on $S^2$ but on the
  path itself. Points $b_1$ and $b_2$ on $S^3$ have the same image B
  on $S^2$ under the first Hopf map. }
\label{diracfig}
\end{figure*}

This simple calculation demonstrates the deep relation between the
first Hopf map, Dirac's monopole and the Berry's phase. We can think of
the first Hopf map as defining a relation between a $SO(3)$ vector
$n_a$ and a $SO(3)$ spinor $(z_1,z_2)$. This relation is invariant
under a $U(1)$ gauge transformation 
$z_\alpha \rightarrow e^{i\phi} z_\alpha$. If the $n_a$ vector goes
from point $A$ to point $B$ on $S^2$ via two different paths, the
$z_\alpha$ vector is transported from point $a$ to {\it two different
points} $b_1$ and $b_2$ on $S^3$. (See figure \ref{diracfig}). 
The transport
of $z_\alpha$ is uniquely defined by the non-singular gauge potential
$\Omega$. Since the two points $b_1$ and $b_2$ differ from each other
by a pure phase, they are projected to the same point $B$ on $S^2$.
Their phase difference is exactly the Berry's phase. Since $S^3$ is
locally isomorphic to $S^2\times U(1)$, it encapsulates the full quantum
information, including both the direction of the $SO(3)$ vector and
the phase of the $SO(3)$ spinor. This way, we can also
understand the singularity associated with a Dirac monopole as 
arising from the projection from a non-singular $U(1)$ gauge field on
$S^3$.

There is an elegant generalization from these considerations 
to the case of a $SO(5)$ spinor. The $SU(2)$ holonomy of a $SO(5)$
spinor and the Yang monopole are deeply related to the second Hopf
map from $S^7$ to $S^4$ and a non-commutative division algebra 
called quaternions. A quaternion with three
imaginary units ${\bf i}$, ${\bf j}$ and ${\bf k}$ is a generalization
of a complex number with one imaginary unit $i$. These three
imaginary units mutually {\it anticommute} and they
obey the following multiplication table:
\begin{eqnarray}
{\bf i}^2 = {\bf j}^2 = {\bf k}^2 = -1 \ \ ; \ \
{\bf i} {\bf j} = {\bf k}\ \ ; \ \ {\bf j} {\bf k} = {\bf i}\ \ ; \ \
{\bf k} {\bf i} = {\bf j}
\end{eqnarray} 
It is simple to understand quaternions from a physicist's view
point, since the algebra of these three imaginary units 
${\bf\Sigma}^\lambda = ({\bf i,j,k})$ is identical to the algebra of three
Pauli matrices $\sigma^\lambda$, and a one-to-one correspondence between them
is therefore possible. A quaternion can be expressed as
\begin{eqnarray}
{\bf q} = q^0  + q^1~ {\bf i} + q^2~ {\bf j} + q^3~ {\bf k}
\end{eqnarray} 
and therefore has four real components.
Complex conjugation of
quaternions is defined as the operation that changes the sign of the
imaginary parts, i.e. of the last three components. Magnitude of the
quaternion is given by $| {\bf q} |^2 = {\bf q}^{*} {\bf q}$.
Using a two component 
quaternion, we can parameterize the $S^7$ sphere by
\begin{eqnarray}
{\bf q}_\alpha = \left( \begin{array}{c} {\bf q}_1 \\ {\bf q}_2 
\end{array} \right) \ \ ; \ \
{\bf q}^{\dagger} {\bf q} = | {\bf q}_1 | ^2 + | {\bf q}_2 |^2 = 1
\label{quaternion}
\end{eqnarray} 
With this notation, the second Hopf map can be expressed as a
simple generalization of ({\ref{eta1}):
\begin{eqnarray} 
\eta_2:~n_a = {\bf q}^{\dagger} \gamma_a {\bf q} 
\label{eta2}
\end{eqnarray}
Here $n_a$ is a five dimensional {\it real} vector, and the five
$2\times 2$ quaternionic valued $\gamma$ matrices are defined as
a simple generalization of the Pauli matrices 
\begin{eqnarray}
\gamma_1 = \left( \begin{array}{cc} 0 & 1 \\ 1 & 0 \end{array} \right)
\hspace{1cm} 
\gamma_2 = \left( \begin{array}{cc} 0 & -{\bf i} \\ {\bf i} & 0
  \end{array} \right) 
\hspace{1cm} 
\gamma_3 = \left( \begin{array}{cc} 0 & -{\bf j} \\ {\bf j} & 0
  \end{array} \right) 
\nonumber\\
\gamma_4 = \left( \begin{array}{cc} 0 & -{\bf k} \\ {\bf k} & 0
  \end{array} \right) 
\hspace{1cm}
\gamma_5 = \left( \begin{array}{cc} 1 & 0 \\ 0 & - 1 \end{array} \right)
\hspace{3.75cm}
\end{eqnarray}
From the norm of ${\bf q}_\alpha$ defined in (\ref{quaternion}) and the
properties of the $\gamma$ matrices, one can easily show that
$n_a^2=1$. Therefore, (\ref{eta2}) defines a mapping from $S^7$ to
$S^4$.
It may be compared with Section \ref{AdEvol} where an $SO(5)$ spinor was
represented by a four component complex vector $\Psi_{\alpha}$ and equation
(\ref{spinorH}) implied that each spinor was mapped into an $SO(5)$
vector by 
\begin{eqnarray} 
n_a = \Psi^{\dagger} \Gamma_a \Psi 
\label{n-a}
\end{eqnarray}
Connection between two representations is easily established
by choosing
\begin{eqnarray}
\Psi^{T} = \{ { q}_2^1 + i { q}_2^2,
                 { q}_1^1 + i { q}_1^2,
                 { q}_2^0 - i { q}_2^3,
                 { q}_1^0 - i { q}_1^3 
\} 
\label{spinor-quaternion}
\end{eqnarray}
which makes  (\ref{eta2}) and (\ref{n-a}) 
identical, and turns the second equation in (\ref{quaternion}) into a
normalization condition for the spinor wavefunction.

Any $SU(2)$ gauge field ${\cal A}$ can be expressed 
in terms of the Pauli matrices as 
${\cal A}^a = {\cal A}^{a\lambda} \sigma_\lambda$. Because of the isomorphism
between the Pauli matrices $\sigma^\lambda$ 
and the three imaginary quaternionic units
${\bf\Sigma}^\lambda = ({\bf i,j,k})$, a $SU(2)$ gauge field can also
be expressed as a imaginary quaternionic field:
\begin{eqnarray} 
i~ {\bf A}^a = A^{\lambda a} {\bf\Sigma}_\lambda
\label{imquat}
\end{eqnarray}
Using this observation and directly substituting the definition of
the second Hopf map (\ref{eta2}) into the $SU(2)$ gauge field of a 
Yang monopole (\ref{calA+}), one finds that it induces a $SU(2)$ gauge
field on $S^7$ defined by 
\begin{eqnarray}
{\bf\omega} = i {\cal A}^a dn_a = {\bf\omega}^\dagger_\alpha d{\bf q}_\alpha + 
{\bf\omega}_\alpha d{\bf q}^\dagger_\alpha  
= -\frac{1}{2 |{\bf q}_1|^2}
[{\bf q}_1 ({\bf q}_2^\dagger d{\bf q}_2 - d{\bf q}^\dagger_2 {\bf q}_2)
{\bf q}^\dagger_1 - |{\bf q}_2|^2
(d{\bf q}_1 {\bf q}_1^\dagger - {\bf q}_1 d{\bf q}^\dagger_1)]
\label{SU2-s}
\end{eqnarray}
where we established
  correspondence between quaternions and Pauli matrices as 
\begin{eqnarray}
{\bf i} \rightarrow i \sigma_y \hspace{1cm} 
{\bf j} \rightarrow - i \sigma_x \hspace{1cm} 
{\bf k} \rightarrow - i \sigma_z
\label{quaternion-Pauli}
\end{eqnarray}

The $SU(2)$ gauge field in (\ref{SU2-s}) is singular on a three
dimensional ``great 
sphere" $|{\bf q}_1|=0$ of $S^7$. Fortunately, just like the case of
the first Hopf map, the singularity can be completely removed by a 
$SU(2)$ gauge transformation defined by a unitary quaternion
${\bf u}={\bf q}_1/|{\bf q}_1|$. The resulting $SU(2)$ gauge field
\begin{eqnarray}
{\bf \Omega} = {\bf u}^{-1} {\bf \omega} {\bf u} + {\bf u}^{-1} d{\bf
  u} = -\frac{1}{2}~ ({\bf q}_\alpha^\dagger d{\bf q}_\alpha 
- d{\bf q}^\dagger_\alpha {\bf q}_\alpha)
\label{Omega-form}  
\end{eqnarray} 
This $SU(2)$ gauge field over $S^7$ appears as a direct quaternionic
generalization of the $U(1)$ gauge field over $S^3$, and it is 
non-singular everywhere. Similarly, one can map the singular $SU(2)$
gauge field $\tilde{\cal A}_+$ on $S^3$ to a singular $SU(2)$ gauge
field $\tilde{\bf \omega}$ on $S^7$. Upon removing the singularity 
by a unitary quaternion ${\tilde{\bf u}}={\bf q}_2/|{\bf q}_2|$,
the resulting $SU(2)$ gauge field is again given by the non-singular
gauge field ${\bf \Omega}$.

Like the previous calculation, this calculation demonstrates the 
deep relation between the second
Hopf map, Yang's monopole and the $SU(2)$ holonomy of a $SO(5)$
spinor. 
We can think of
the second Hopf map as defining a relation between a $SO(5)$ vector
$n_a$ and a normalized $SO(5)$ spinor ${\bf q}_\alpha = ({\bf q}_1,{\bf q}_2)$.
This relation is invariant under a quaternionic unitary transformation
${\bf q}_\alpha \rightarrow {\bf q}_\alpha {\bf u}$ with 
${\bf u}^\dagger {\bf u} = 1$. Such a quaternionic unitary transformation
is identical to a $SU(2)$ gauge transformation. In this case,
if the $n_a$ vector goes
from point $A$ to point $B$ on $S^4$ via two different paths, the
${\bf q}_\alpha$ vector is transported from point $a$ to 
{\it two different
points} $b_1$ and $b_2$ on $S^7$. (See figure \ref{yangfig}).
\begin{figure*}[h]
\centerline{\epsfysize=3.8cm 
\epsfbox{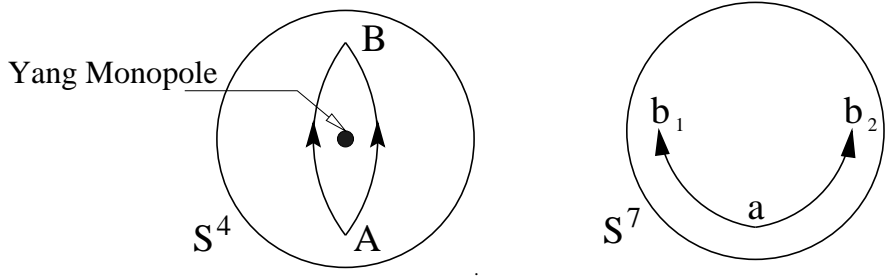}
}
\caption{When a vector is transported on $S^4$ along two different
  paths a spinor is 
  transported on $S^7$ to two different final states $b_1$ and $b_2$,
  which have the same image B on $S^4$.  }
\label{yangfig}
\end{figure*}
 The transport
of ${\bf q}_\alpha$ is uniquely defined by the non-singular gauge potential
${\bf \Omega}$. Since the two points $b_1$ and $b_2$ differ from each 
other by a $SU(2)$ gauge transformation, 
they are projected to the same point $B$ on $S^4$.
Their difference is exactly the $SU(2)$ holonomy of a $SO(5)$ spinor. 
Since $S^7$ is
locally isomorphic to $S^4\times SU(2)$, it encapsulates the full quantum
information, including both the direction of the $SO(5)$ vector and
the $SU(2)$ holonomy of a $SO(5)$ spinor.

Another way of seeing the invariance of the second
  Hopf map under $SU(2)$ gauge transformation comes from returning to
  a representation of the $SO(5)$ spinor as  a four
  component complex vector $\Psi$. Choosing components of $\Psi$
  as 
\begin{eqnarray}
\Psi = \{ \psi_1, \phi_1,
  \psi_2^{\dagger},\phi_2^{\dagger} \}^{T}
\end{eqnarray} 
gives for the second Hopf map  $n_a = \Psi^{\dagger} \Gamma_a \Psi $
\begin{eqnarray}
n_1 &=& \phi^{\dagger}_{\alpha} \psi_{\alpha} + \psi^{\dagger}_{\alpha}
\phi_{\alpha} \nonumber\\    
n_2 &=& i( \phi^{\dagger}_{\alpha} \psi_{\alpha} - \psi^{\dagger}_{\alpha}
\phi_{\alpha} ) \nonumber\\    
n_3 &=& \epsilon_{\alpha \beta} ( \phi_{\alpha} \psi_{\beta} +
\psi^{\dagger}_{\alpha} \phi_{\beta}^{\dagger} ) \nonumber\\    
n_4 &=& i \epsilon_{\alpha \beta} ( \phi_{\alpha} \psi_{\beta} -
\psi^{\dagger}_{\alpha} \phi_{\beta}^{\dagger} ) \nonumber\\    
n_5 &=& \phi^{\dagger}_{\alpha} \phi_{\alpha} - \psi^{\dagger}_{\alpha}
\psi_{\alpha} 
\end{eqnarray} 
One can easily see that this exhausts all possible bilinear
combinations  that are both real and invariant under simultaneous $SU(2)$
rotations of $\phi$ and $\psi$ \cite{thanks}.

To conclude this section we remark that analysis
presented here leads us to an intriguing conclusion that not all
spheres are created equal. $S^3$ and $S^7$ are 
endowed with 
a unique algebraic structure and associated non-singular complex 
(\ref{Omega-eq}) and quaternionic (\ref{Omega-form}) gauge connections.
This deep connection between the algebra and geometry underlies the 
non-trivial holonomy of  $SO(3)$ and  $SO(5)$ spinors.

\subsection{ $HP(1)$ Representation of the SO(5) Nonlinear $\sigma$ Model
  }

It has long been known that the $SO(3)$ nonlinear sigma model used to
describe the low energy physics of magnetic systems has an
interesting representation due to the existence 
of the first Hopf map: $CP(1)$ representation
\cite{Polyakov}.  Indeed one observes that when we define the first
Hopf map as in equations (\ref{znorm}) and (\ref{eta1}) of Section
\ref{HopfMap}, 
the partition function for the nonlinear sigma model may be written as
a path integral over $z$ and gauge field $A$
\begin{eqnarray}
Z= \int {\cal D} n e^{ - \int_0^\beta  d \tau \int d^d x \frac{1}{4} (
  \partial_{\mu} n^a )^2 } = \int {\cal D} A {\cal D} z  e^{ - \int_0^\beta  d
  \tau \int d^d x~ | ( \partial_{\mu} - i A_{\mu}
) z |^2 }
\label{partitionNLSM}
\end{eqnarray}
This may be easily proven by noticing that the action on the right
hand side of
(\ref{partitionNLSM}) is quadratic in $A$, so integration over the
gauge field amounts to taking the saddle point value
\begin{eqnarray}
A_{\mu} =  - \frac{i}{2} \left[ z^{\dagger} \partial_{\mu} z - (
  \partial_{\mu} z^{\dagger}) z \right]
\label{Aeq}
\end{eqnarray} 
After that simple algebra gives
\begin{eqnarray}
\frac{1}{4} ( \partial_{\mu} n^a )^2 = | ( \partial_{\mu} - i A_{\mu}
) z |^2 
\label{CP1-eqn}
\end{eqnarray}
which proves equation (\ref{partitionNLSM}).
Notice that the gauge
field in (\ref{Aeq}) is the same as in (\ref{Omega-eq}), which is not
surprising since the origin of both gauge fields is the invariance of
(\ref{eta1}) and (\ref{CP1-eqn}) under gauge
transformations 
\begin{eqnarray}
 z_i & \rightarrow & e^{i \phi} z_i \nonumber\\
 A_{\mu} & \rightarrow & A_{\mu} + \partial_{\mu} \phi 
\label{u1-gt}
\end{eqnarray}

Analogously one can prove that the second Hopf map, defined by
(\ref{quaternion}) and (\ref{eta2}) in Section \ref{HopfMap},
gives rise to an $HP(1)$ representation of the $SO(5)$ nonlinear sigma
model 
\begin{eqnarray}
Z= \int {\cal D} n e^{ - \int_0^\beta  d \tau \int d^d x \frac{1}{4} (
  \partial_{\mu} n^a )^2 } &=& \int {\cal D} {\bf B} {\cal D} {\bf q}
  e^{ - \int_0^\beta  d 
  \tau \int d^d x~  (D_{\mu} {\bf q})^{\dagger} (D_{\mu} {\bf q}) }
  \nonumber\\ 
D_{\mu}{\bf  q} &=& \partial_{\mu} {\bf q} -  {\bf q} {\bf B}_{\mu}
\label{partitionQNLSM}
\end{eqnarray}
where similarly to  $CP(1)$, the $HP(1)$ action
\begin{eqnarray}
\frac{1}{4} (\partial_{\mu} n^a )^2 = (D_{\mu} {\bf q})^{\dagger}
(D_{\mu} {\bf q}) 
\label{HPaction}
\end{eqnarray}
as well as the second Hopf map (\ref{eta2}) are invariant under 
gauge transformation 
\begin{eqnarray}
{\bf q}_{\alpha} & \rightarrow & {\bf q}_{\alpha} {\bf u} 
\nonumber\\
{\bf B}_{\mu} & \rightarrow & {\bf u}^{-1} {\bf B}_{\mu} {\bf u} +
{\bf u}^{-1} \partial_{\mu}  {\bf u} 
\label{su2-gt}
\end{eqnarray}  
with ${\bf u}$ being any quaternion that satisfies ${\bf u}^{\dagger}
{\bf u} =1 $. There is, however, an important difference between
gauge transformations in (\ref{su2-gt}) and (\ref{u1-gt}). In the
former case they are non-Abelian $SU(2)$ gauge transformations whereas
in the latter case they are Abelian $U(1)$ gauge transformations. 

It is also not surprising that the saddle point value of the 
gauge field ${\bf B}$ in (\ref{partitionQNLSM}) is given by 
\begin{eqnarray}
{\bf B}_{\mu}  = -\frac{1}{2}~ [~ {\bf q}_\alpha^\dagger
  \partial_{\mu} {\bf q}_\alpha  
- ( \partial_{\mu}{\bf q}^\dagger_\alpha )~ {\bf q}_\alpha~ ]
\label{B-form}  
\end{eqnarray} 
which agrees exactly with expression (\ref{Omega-form}) that we obtained in the
study of the $SO(5)$ spinor holonomy (notice that both ${\bf B}$ and
the one form ${\bf \Omega}$ in (\ref{Omega-form}) are quaternionic
imaginary, or anti-hermitian). 

So far the $HP(1)$ model is only an interesting reformulation of the
$SO(5)$ nonlinear $\sigma$ model. However, it has the advantage that
possible topological terms can be easily identified within this formulation.
In $2+1$ dimension, one could add the second Hopf invariant, or the
non-Abelian Chern-Simons term
\begin{eqnarray}
\frac{n}{8 \pi^2}~  Tr \int d^3 x~ \epsilon^{\mu\nu\lambda} \left(
{\bf B}_{\mu}  \partial_{\nu} {\bf B}_{\lambda}
 - \frac{2  }{3} {\bf B}_{\mu} {\bf B}_{\nu} {\bf B}_{\lambda} \right)
\end{eqnarray} 
to the $HP(1)$ action (\ref{partitionQNLSM}), where $Tr$ implies
taking the real part of a quaternion ( in representation where
quaternions are $2\times2$ matrices as in (\ref{quaternion-Pauli}), it
corresponds to actually taking a trace ). 
Because the Chern-Simons term changes by 
\begin{eqnarray}
\frac{n}{16 \pi^2}~  Tr \int d^3 x~  \epsilon^{\mu\nu\lambda}  \left(
   {\bf u}^{-1} ( \partial_{\mu} 
   {\bf u} ) {\bf u}^{-1} ( \partial_{\nu} {\bf u} ) {\bf u}^{-1}
   ( \partial_{\lambda} {\bf u} ) \right) 
\label{CSexp}
\end{eqnarray} 
under global gauge transformation (\ref{su2-gt}), the coupling
constant $n$ has to be quantized  ( for $n=1$ expression in
(\ref{CSexp}) is a winding number of the gauge transformation which is
always integer), i.e. only integer values of $n$ are allowed. 
In
$3+1$ dimension, one could 
add a $\theta$ term,
\begin{eqnarray}
& \theta & Tr \int d^4 x~  \epsilon^{\mu\nu\lambda\sigma} {\bf F}_{\mu\nu}
{\bf F}_{\lambda\sigma} \nonumber\\
{\bf F}_{\mu\nu} &=& \partial_{\mu} {\bf B}_{\nu} - \partial_{\nu} {\bf
  B}_{\mu} + [ {\bf B}_{\mu}, {\bf B}_{\nu} ] 
\end{eqnarray} 
to the action $HP(1)$ action (\ref{partitionQNLSM}).
In $6+1$ dimensions, Wu and Zee \cite{WuZee} showed that one could
also add an Abelian Chern-Simons term for the 3 index gauge field
introduced in (\ref{threeindex}).
However, we haven't
yet identified microscopic models which would give rise to these topological
terms in the action.

\subsection{ Wigner - von Neumann Class } 

The non-trivial holonomy of the quantum mechanical wave functions are 
intimately related to the degeneracy point of the Hamiltonian. Wigner and
von Neumann classified the degeneracy points into three classes
according to the {\it generic
symmetries} of the Hamiltonian. Time-reversal invariant systems without
Kramers degeneracy (system of bosons or even number of fermions) belong
to the orthogonal class. The generic degeneracy point has co-dimension
$2$, {\it i.e.} one needs to tune two parameters simultaneously to zero
in order to reach a point of degeneracy. Time-reversal breaking system
belong to the unitary class, and the degeneracy point has codimension
$3$. Berry's phase of a $SO(3)$ spinor describes the non-trivial holonomy
around this point singularity in the $3$ dimensional parameter space,
which can be described  as a Dirac monopole. Time-reversal invariant
system with 
Kramers degeneracy (e.g. system with odd number of fermions) belong
to symplectic class. The degeneracy point for this class has codimension
$5$. Unlike the two previous classes, we have to consider
a $4\times 4$ matrix problem in this case, in order to describe the
level crossings of two different Kramers doublets. In his seminal work
on random matrix theory, Dyson \cite{Dyson} noticed that this
situation can be best 
described as a $2\times 2$ quaternionic matrix problem. In a thorough 
analysis, Avron {\it et al} \cite{Avron1,Avron2} showed that the natural 
symmetry for the symplectic class is therefore a unitary rotation of 
a two component quaternion, which is isomorphic to a $SO(5)$ symmetry
group. The non-trivial holonomy of a $SO(5)$ spinor arises when one
encircles this degeneracy point in $5$ dimensional space, which acts
as a Yang monopole.    

In this context, it is highly remarkable that $SO(5)$ symmetry emerges
naturally in a {\it generic} fermionic system with time reversal symmetry.
To make the above discussion more explicit, let us recall that close
to the degeneracy point, a generic Hamiltonian without Kramers degeneracy can be
expressed as a $2\times 2$ Hermitian matrix
\begin{eqnarray}
{\cal H} = \left( \begin{array}{cc} a & d \\ c & b \end{array} \right)
\hspace{2cm} a,b \in {\cal R},~ d^* = c \in {\cal C}
\label{H-degeneracy}
\end{eqnarray}
whose eigenvalues are 
\begin{eqnarray}
\lambda_{\pm} = \frac{
a+b \pm \sqrt{(a+b)^2 -4 ( ab - | d |^2 )}
}{2} 
\end{eqnarray}
Degeneracy requires
$
(a -b)^2 + 4 | d |^2 = 0
$,
which is only satisfied when $a=b$ and $Re~d = Im~d = 0$. When the
Hamiltonian is real ( for 
systems without time-reversal symmetry breaking ) we have only
two conditions for the degeneracy point, i.e. the co-dimension is
$2$. But when  the time reversal-breaking is present, we have three
conditions and the codimension is $3$.

Up to a overall additive constant, this $2\times 2$ 
Hamiltonian can be mapped onto 
a effective spin $1/2$ problem in a magnetic field:
${\cal H}_{\alpha\beta} = n_a \sigma^a_{\alpha \beta}$, 
where $n_a$ labels the three directions away from the degeneracy
point $n_a=0$. 
Wave functions defined for parameters on $S^2$
enclosing the singularity are non-degenerate, but experience
non-trivial holonomy due to the degeneracy point, as discussed in
Section \ref{AdEvol}.

Close to the degeneracy point of two Kramers doublets, the
Hamiltonian is a $4\times 4$ matrix, and can be represented by the
matrix of a spin $3/2$ particle. However, because of the time
reversal symmetry, the Hamiltonian has to be quadratic in these
spin $3/2$ matrices. Avron {\it et al} \cite{Avron1,Avron2} showed
that a {\it generic} 
Hamiltonian close to the degeneracy point can be expressed as
\begin{eqnarray}
{\cal H}_{\alpha\beta} = \sum_{i,j} Q_{ij} S^i_{\alpha\gamma}S^j_{\gamma\beta}
\label{quadrupole}  
\end{eqnarray}
where $S^i_{\alpha\gamma}$ are the three $4\times 4$ spin $3/2$ matrices,
and $Q_{ij}$ is a $3\times 3$ {\it real symmetric traceless} matrix. 
Since $Q_{ij}$ has $5$ real entries, and one needs to tune all of them
to zero to get a degeneracy of two Kramers doublets, therefore the
codimension of the 
degeneracy point is $5$. In fact,
this is nothing but a quadrupole spin Hamiltonian first used by Zee
\cite{zee} and other authors\cite{pines,arovas} to 
illustrate the concept of non-Abelian holonomy.

This generic Hamiltonian for the symplectic class
(\ref{quadrupole}) has a hidden $SO(5)$ symmetry. Since $Q_{ij}$ has 
exactly five entries, we can choose a orthonormal basis set 
$Q^a_{ij}$, $(a=1,..5)$ satisfying $Tr(Q^a Q^b)=\frac{2}{3}\delta^{ab}$
and expand $Q_{ij}$ as $Q_{ij}=\sum_a n_a Q^a_{ij}$. In this
representation, the generic Hamiltonian (\ref{quadrupole}) 
can be expressed as 
\begin{eqnarray}
{\cal H}_{\alpha\beta} = n_a \Gamma^a_{\alpha\beta}\ \ ; \ \
\Gamma^a_{\alpha\beta} = S^i_{\alpha\gamma} Q^a_{ij} S^j_{\gamma\beta} 
\end{eqnarray}
At this point, the hidden $SO(5)$ symmetry becomes manifest.
Since the five $\Gamma$ matrices obey a $SO(5)$ Clifford algebra, this
Hamiltonian is identical to the $SO(5)$ spinor Hamiltonian (\ref{spinorH})
discussed in section \ref{AdEvol}.

Therefore, we see that $SO(5)$ symmetry arises naturally close to {\it any generic}
degeneracy point of a time reversal invariant 
Hamiltonian with Kramers degeneracy. Both 
AF and dSC states in the high $T_c$ problem are time reversal invariant,
and their quasi-particle states form degenerate pairs. At the
fermi liquid point, two of these pairs become additionally degenerate. 
Therefore $SO(5)$ symmetry emerges naturally in this system. 
The non-Abelian holonomy of the SDW and BCS quasi-particles arises close to
the fermi liquid degeneracy point, which is identified with a Yang 
monopole in the parameter space.

\section{ Effective Adiabatic Action for the SO(5) Non-Linear $\sigma$ Model }

\subsection{Effective Action Arising from Berry's Phase}
\label{BerrysPhase}

Let us now consider the problem of a rigid $SO(5)$ rotor 
interacting with a $SO(5)$ spinor. We want to integrate out the 
spinor degrees of freedom and ask whether this procedure will generate
any additional contributions to the action for the $SO(5)$
rotator ( see  Kuratsuji and Iida \cite{KuratsujiIida} and M.~Stone
\cite{Stone} for such discussion in the case of coupled $SO(3)$ spinor
and rotor ).

Our system is described by the Hamiltonian
\begin{eqnarray}
{\cal H} &=& {\cal H}_0 + {\cal H}_I \\
{\cal H}_0 &=& \frac{1}{2I} \sum_{a<b} L_{ab}^2  
\label{H-I}
\end{eqnarray}
Here ${\cal H}_0 $ is the usual action of the $SO(5)$ rigid rotor
\cite{science} 
with $L_{ab}$ being the generators of rotation, and  ${\cal H}_I$ is
the Hamiltonian (\ref{spinorH}). 
The states of this composite system are defined as product states 
\begin{eqnarray}
| \Psi_{\alpha} [n],n_a \rangle = | \Psi_{\alpha} [n] \rangle \otimes
| n_a \rangle  
\end{eqnarray}
where $ | n_a \rangle $ are the states of the $SO(5)$ vector and $|
\Psi_{\alpha} [n] \rangle $ are spinor states. We choose $|
\Psi_{\alpha} [n] \rangle $ to be eigenstates of ${\cal H}_I
(n,\Psi)$, i.e. they diagonalize (\ref{H-I}) for a given orientation
of $n_a$.
\begin{eqnarray}
{\cal H}_I(n,\Psi) | \Psi_{\alpha} [n] \rangle = E_{\alpha} |
\Psi_{\alpha} [n] \rangle 
\end{eqnarray}

The partition function is defined as
\begin{eqnarray}
Z(\beta) = \sum_{n,\alpha} \langle \Psi_{\alpha} [n(0)],n_a(0) | e^{ - \beta
  {\cal H}  } |
| \Psi_{\alpha} [n(0)],n_a(0) \rangle 
\end{eqnarray}
with summation going over all states of $n_a$ and $\Psi_{\alpha}$. Let
us perform time 
discretization and insert resolution of identity for the
$n$-states. We make $N$ intervals of length $\epsilon$ with
$N\epsilon=\beta$: 
\begin{eqnarray}
Z(\beta) = \prod_{k=1}^{N-1} \int d\mu(n_k) \sum_{n(0),\alpha} 
\langle \Psi_{\alpha} [n(0)],n_a(0) | e^{ - {\cal H} \epsilon } |
n(N-1) \rangle  
\langle  n(N-1) | \dots | \Psi_{\alpha} [n(0)],n_a(0) \rangle 
\end{eqnarray}
where $d \mu(n_k)$ is an invariant measure of integration for the $|
n_a \rangle $
states at interval $k$.
When $\epsilon$ is small we have 
\begin{eqnarray}
\langle n(k)| e^{- {\cal H} \epsilon } | n(k-1) \rangle = 
\langle n(k)| e^{- {\cal H}_0 \epsilon } | n(k-1) \rangle
e^{-  {\cal H}_I(n(k)) \epsilon }
\end{eqnarray}
and 
\begin{eqnarray}
Z(\beta) = &&\sum_{n(0),\alpha} \langle n(0) | e^{- {\cal H}_0 \epsilon } |
n(N-1) \rangle \langle n(N-1) |  e^{- {\cal H}_0 \epsilon } |
n(N-2) \rangle \dots \langle n(1) |  e^{- {\cal H}_0 \epsilon } |
n(0) \rangle \times \nonumber\\ \times
&&\langle \Psi_{\alpha} [n(0)] | e^{ - {\cal H}_I(n_N) \epsilon}e^{
  - {\cal H}_I(n_{N-1}) 
  \epsilon} \dots  e^{ - {\cal H}_I(n_1) \epsilon} | \Psi_{\alpha} [n(0)]
\rangle
\label{KT}
\end{eqnarray}
The first line of the last equation is clearly the partition function
of the $SO(5)$ rotor. Following Auerbach \cite{Auerbach} we can write it as
\begin{eqnarray}
&&\langle n(k) | e^{- {\cal H}_0 \epsilon } | n(k-1) \rangle = \nonumber\\
&=&\int d \mu(p_k) \langle n(k) | p(k) \rangle \langle p(k) | e^{- {\cal
    H}_0 \epsilon } | n(k-1) \rangle \nonumber\\
&=&  \int d \mu(p_k)~ exp\{ i p_{a,k} (
  n_{a,k} - n_{a,k-1} )  -  \frac{1}{2I} \sum_{a<b} ( p_a n_b - p_b
n_a)^2 \epsilon    \} \nonumber\\
&=& const \times exp\{- \frac{(n_a(k)-n_a(k-1))^2}{2 I \epsilon^2}
\epsilon \}  
\label{K0}
\end{eqnarray}
In the second line of equation (\ref{KT}) we insert the resolution of
identity at each $\tau$ step 
$\sum_{\beta_j} | \Psi_{\beta_j} [n_j] \rangle \langle \Psi_{\beta_j}
[n_j] | =1 $, then
\begin{eqnarray}
T_{n\alpha;n\alpha} = \langle \Psi_{\alpha} [n(0)] | e^{ - h(n_N)
  \epsilon}e^{ - h(n_{N-1}) 
  \epsilon} \dots  e^{ - h(n_1) \epsilon} | \Psi_{\alpha} [n(0)]
\rangle = \nonumber\\
\prod_{j=1}^{N-1} \sum_{\beta_j} \langle \Psi_{\alpha} [n(0)] | e^{ -
  h(n_N) } | \Psi_{b_{N-1}} 
  [ n_{N-1}] \rangle \langle  \Psi_{b_{N-1}} [ n_{N-1}] | \dots |
  \Psi_{\alpha} [n(0)] \rangle
\end{eqnarray}
We now recall the adiabatic hypothesis and realize that in summation
over intermediate spinor states in the last expression we only have to
consider two states of the 
same energy as the initial state ( recall discussion of double
degeneracy of spinor states in section
\ref{AdEvol} ). In the limit $\epsilon \rightarrow
0$ the last expression becomes the path ordered exponential 
\begin{eqnarray}
T_{n\alpha;n\alpha} = e^{ - \beta E_{\alpha} } \left( T_{\tau} exp\{ i~\oint
{\cal A}^a [n(\tau)] dn_a \} \right)_{\alpha\alpha}
\label{Tspinor}
\end{eqnarray}
Here 
\begin{eqnarray}
{\cal A}^a_{\alpha\beta} [n(t)] = \frac{1}{i} \langle \Psi_{\alpha}[n] |
\frac{\partial}{\partial n_a} | \Psi_{\beta}[n] \rangle
\label{BP}
\end{eqnarray}
and indices $\alpha$, $\beta$ take two values that correspond to the two
degenerate states of the same energy as the initial state, but with
the new orientation of the $n$-field ( new instantaneous eigenstates in the
language of section \ref{AdEvol} ). Therefore we arrive at the same $SU(2)$
gauge connection ${\cal A}^a$ as defined in equations
(\ref{A-definition}) and (\ref{A-definition}). By combining equations
(\ref{KT}) and 
(\ref{Tspinor}) we see that this gauge field ${\cal A}^a$ defines a
non-trivial  
non-Abelian topological term in the action for the $SO(5)$ rotor
\begin{eqnarray}
Z(\beta) = \int {\cal D} n_{\tau} Tr \left[ exp \left\{ - \oint_0^{\beta}
      \left(\frac{d 
      n_a}{d \tau} \right)^2 d\tau + i \oint {\cal A}^a [n(\tau)] d n_a
      \right\} \right]
\label{Knonabelian}
\end{eqnarray}

In Chapter \ref{HopfMap} we saw how the first term in this path
integral may be written using two quaternions ${\bf q} = ( {\bf q}_1,
{\bf q}_2)^{T}$ 
\begin{eqnarray}
\frac{1}{4} \left( \frac{d n_a}{d\tau}\right)^2 =  ( {\cal D}_{\tau}
{\bf q} )^{\dagger} 
( {\cal D}_{\tau} {\bf q} )  \nonumber\\ 
{\cal D}_{\tau} {\bf q} =~ \partial_{\tau} {\bf q} - {\bf q} {\bf B} ({\tau})
\hspace{0.5cm} & & \hspace{0.5cm} 
{\bf B} (\tau) =  - \frac{1}{2} \left[ {\bf q}_{\alpha}^{\dagger}
  \partial_{\tau} 
{\bf q}_{\alpha} - ( \partial_ {\tau}{\bf q}_{\alpha}^{\dagger}) {\bf
  q}_{\alpha} \right]
\end{eqnarray}
As it turns out, the second term of equation (\ref{Knonabelian}) can
also be represented using quaternions 
\begin{eqnarray}
i {\cal A}^a [n(\tau)] d n_a \rightarrow {\bf
  q}_{\alpha}^{\dagger} \partial_{\tau}  
{\bf q}_{\alpha} d\tau = - {\bf B}({\tau}) d\tau
\label{quatBP}
\end{eqnarray}
Therefore, the $HP(1)$ representation of (\ref{Knonabelian}) reads:
\begin{eqnarray}
Z(\beta) = \int {\cal D} {\bf B} {\cal D} {\bf q}~ Tr \left[
  exp \left\{ - \int_0^\beta  d\tau~ \left( (D_{\tau} {\bf q})^{\dagger}
  (D_{\tau} {\bf q})  + 
  {\bf B}({\tau})  \right) \right\} \right]
\end{eqnarray}

\section{Physical Interpretation of the SU(2) Holonomy}
\label{holonomysection}

\subsection{Holonomy on a Rung in the $SO(5)$ Ladder Problem }
\label{2sitesection}
Having established the necessary mathematical framework, we are now
in a position to discuss the physical application to the holonomy
between SDW and BCS quasi-particles. As a warm-up exercise, let us
first consider a two site problem ( see Figure \ref{twosite}).
\begin{figure*}[h]
\centerline{\epsfysize=3.8cm 
\epsfbox{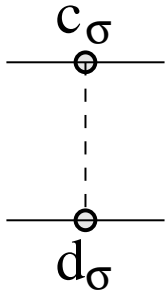}}
\caption{Two site problem. Electrons on the upper site of the of the
  rung are created by $c_{\sigma}^{\dagger}$ operators and
  $d_{\sigma}^{\dagger}$ creates electrons on the lower site.}
\label{twosite}
\end{figure*}
Recently, Scalapino, Hanke and
one of us (SCZ)\cite{szh} studied a $SO(5)$ symmetric ladder
model, and classified the $SO(5)$ operator content for the
two site problem. They introduced a $SO(5)$ spinor
\begin{eqnarray}
\Psi_\alpha = \left(
\begin{array}{c}
c_\sigma\\
d^\dagger_\sigma
\end{array}
\right)
\label{even-spinor}
\end{eqnarray}
where $c_{\sigma}^{\dagger}$ and $d_{\sigma}^{\dagger}$ operators
create electrons on the upper and lower sites of the rung
respectively. If we consider the $SO(5)$ vector interaction on the 
same rung, and introduce a time-dependent Hubbard-Stratonovich 
field $n_a(t)$ to decouple the interaction, we obtain the following
effective fermion problem on a rung: 
\begin{eqnarray}
{\cal H}(t) = \Delta~ n_a(t) \Psi^{\dagger}_{\alpha}
\Gamma^a_{\alpha\beta} \Psi_{\beta}
\label{rungH} 
\end{eqnarray}
Our task is to find the ground state and low energy quasi-particle
excitations for this problem and demonstrate the physical 
interpretation of the non-Abelian holonomy.

This physical problem is identical to the mathematical problem we
posed in section \ref{SpinorRot}. We can transform the problem into the
``rotating frame" by decomposing $\Psi_\alpha$ according to
(\ref{decomposition}), with a $S_{\alpha\beta}$ matrix satisfying
(\ref{S-condition1}) and (\ref{S-condition2}). The resulting 
Hamiltonian (\ref{phiH}) in the ``rotating frame" is time independent
and diagonal in the $\Phi_\alpha$ variable. $\Phi_1$ and $\Phi_3$
have positive energy, while $\Phi_2$ and $\Phi_4$ have negative 
energy. According to the Dirac prescription, we fill the negative
energy states and obtain the ground state $|\Omega\rangle$ in
the rotating frame:
\begin{eqnarray}
\Phi^\dagger_2|\Omega\rangle = \Phi^\dagger_4|\Omega\rangle = 0 \ \ ; \ \
\Phi_1|\Omega\rangle = \Phi_3|\Omega\rangle = 0
\label{phiOmega}
\end{eqnarray}
(Notice that the definition of $|\Omega\rangle$ is different from 
reference \cite{szh}).
The corresponding elementary excitations are given by
\begin{eqnarray}
\Phi^\dagger_1|\Omega\rangle\ \ ; \ \
\Phi^\dagger_3|\Omega\rangle\ \ ; \ \
\Phi_2|\Omega\rangle\ \ ; \ \
\Phi_4|\Omega\rangle
\label{phiexcite}
\end{eqnarray}
with {\it degenerate energy} $+\Delta$. 
If the $n_a$ vector points to $n_4$ direction at $t=0$, the ground
state and four elementary excitations can be explicitly expressed
in terms of electron operators $c_\sigma$ and $d_\sigma$, and
they are depicted in Figure \ref{states}.
\begin{figure*}[h]
\centerline{\epsfysize=5.8cm 
\epsfbox{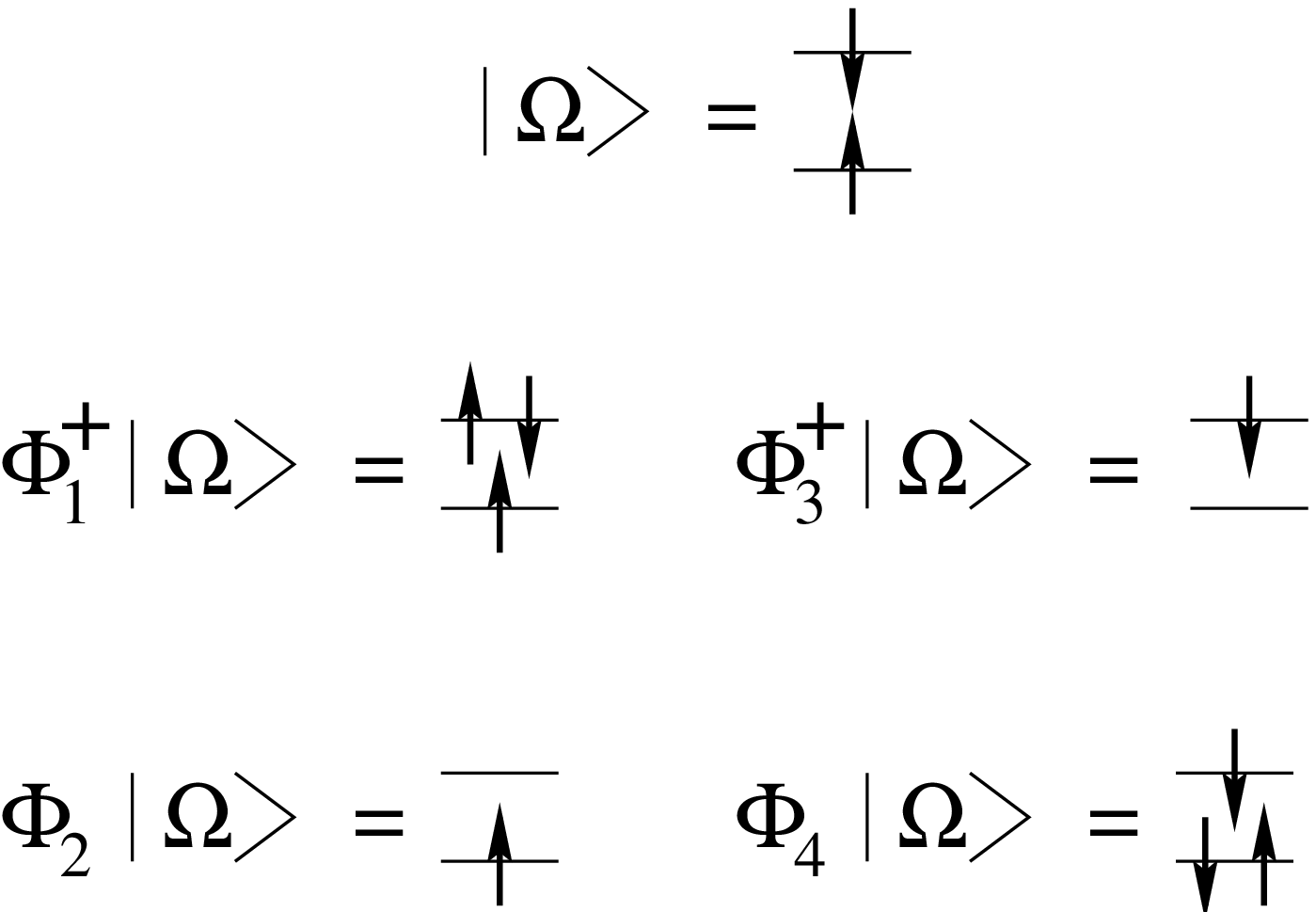}}
\caption{Ground state and single particle excited states on a rung,
  when $n_a$ points in $n_4$ direction.}
\label{states}
\end{figure*}

Within the adiabatic approximation, the $S_{\alpha\beta}$ matrix is
explicitly given by (\ref{Scomb}). The instantaneous ground state 
and elementary excitations in the original frame is simply obtained
from (\ref{phiOmega}) and (\ref{phiexcite}) by substituting 
$\Phi_\alpha=S^\dagger_{\alpha\beta}\Psi_\beta$. Let us now imagine
taking a path of $n_a(t)$ on $S^4$ which starts and returns to $n_4$.
Because of the non-Abelian holonomy discussed in section
\ref{SpinorRot}, the 
$S_{\alpha\beta}$ matrix only returns to itself up to a $SU(2)$
transformation in the $(1,3)$ and $(2,4)$ subspaces. 
In particular, there exist cyclic paths on $S^4$ for which
\begin{eqnarray}
S(t=0) = 
\left( \begin{array}{cccc}
1 & 0 &  0 &  0 \\
0 & 1 &  0 &  0 \\
0 & 0 &  1 &  0 \\
0 & 0 &  0 &  1  \end{array}
\right)\ \ ; \ \ 
S(t=T) = 
\left( \begin{array}{cccc}
0 & 0 &  1 &  0 \\
0 & 0 &  0 &  1 \\
1 & 0 &  0 &  0 \\
0 & 1 &  0 &  0  \end{array}
\right)
\label{S-evolution}
\end{eqnarray}
Under such a cyclic path, the quasi-particle state 
$\Psi^\dagger_1|\Omega\rangle$ interchanges with
$\Psi^\dagger_3|\Omega\rangle$, while the
quasi-particle state 
$\Psi_2|\Omega\rangle$ interchanges with
$\Psi_4|\Omega\rangle$ ( see Figure \ref{evolution}). 
\begin{figure*}[h]
\centerline{\epsfysize=4.0cm 
\epsfbox{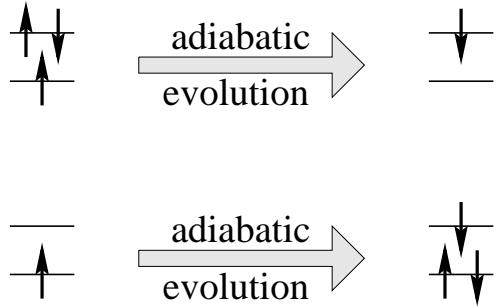}}
\caption{Interchange of degenerate spinor states after cyclic adiabatic
  rotation of the superspin.}
\label{evolution}
\end{figure*}

Therefore, we obtained a simple physical interpretation of the
$SU(2)$ non-Abelian holonomy in terms of interchange of certain degenerate
pairs of quasi-particle states on a rung. It is important to notice that
the degeneracy of these pairs is {\it generic} in any model with
spin rotation and charge conjugation symmetry, and does not require
any special fine tuning of parameters that is required to render the 
model $SO(5)$ symmetric in all sectors\cite{szh}. 
For this reason, we believe
that the non-Abelian holonomy we discussed in this work have general
applicability beyond exactly $SO(5)$ symmetric models.   

We notice that in the course of a cyclic adiabatic evolution,
a SDW quasi-particle can be turned into a SDW quasi-hole (see Figure
\ref{evolution}).  
In this sense, this phenomenon is very analogous to Andreev reflection
at the metal/superconductor interfaces, where a metallic quasi-particle
reflects back as a quasi-hole, and emits a Cooper pair into the 
superconductor. However, close inspection of the spin quantum numbers
shows that a $\pi$ pair \cite{science,prl95}, rather than a Cooper
pair is emitted in our 
problem. Indeed, the states interchanged after cyclic
adiabatic transport of the superspin differ from each other by 
addition/removal of a pair of electrons in a  triplet state ( a
pair of spin up electrons for 
the first pair of degenerate states on figure \ref{evolution}, and a
pair of spin down electrons for 
the second pair of degenerate states on figure \ref{evolution}
). Operators that create such triplet pairs are generators of the
$SO(5)$ symmetry on the rung \cite{szh} ($\pi$ operators), they carry
charge $2$ and spin $1$. 
By contrast, the operator that
produces a $d$-wave Cooper pair creates two
electrons on opposite sites in a  singlet state, therefore it
carries charge 2 and spin 0.


The appearance of the $\pi$ operators as operators of the holonomy is
not surprising and is related to the symmetry content of
Hamiltonian (\ref{rungH}). 
For a fixed direction of $n_a$ the symmetry of this Hamiltonian 
is reduced from $SO(5)$ to $SO(4)$.  For example when $n_a$ points
toward $n_4$, the $SO(4)$ algebra that commutes with the Hamiltonian
is formed 
by operators that rotate between $n_1$, $n_2$, $n_3$, and $n_5$ but
leave $n_4$ invariant. Let $L_{ab}$ be the operators that rotate
between $n_a$ and $n_b$ ( microscopically, such operators are given by
$L_{ab} = \Psi^{\dagger}_{\alpha} \Gamma^{ab}_{\alpha\beta}
\Psi_{\beta}$, where $\Gamma^{ab}= i [\Gamma^{a},\Gamma^{b}]$, see
\cite{prl98-1,szh} for details ), then the generators of the unbroken
$SO(4)$ symmetries are   
\begin{eqnarray}
K_1 = \frac{1}{2}~ ( L_{21} - L_{35} ) &\hspace{1cm}
K_2 = \frac{1}{2}~ ( L_{13} - L_{25} ) &\hspace{1cm}
K_3 = \frac{1}{2}~ ( L_{32} - L_{15} ) \nonumber\\
J_1 = \frac{1}{2}~ ( L_{21} + L_{35} ) &\hspace{1cm}
J_2 = \frac{1}{2}~ ( L_{13} + L_{25} ) &\hspace{1cm}
J_3 = \frac{1}{2}~ ( L_{32} + L_{15} )  \nonumber\\
\left[ K_i, K_j \right] = i~\epsilon_{ijk} K_k     &\hspace{1cm}
\left[ J_i, K_j \right] = 0                        &\hspace{1cm}
\left[ J_i, J_j \right] = i~\epsilon_{ijk} J_k
\label{SO4}
\end{eqnarray}
We have chosen the generators in a form that makes transparent a
property of $SO(4)$ to be factorizable into $SU(2) \times SU(2)$.
The eigenstates of our Hamiltonian have to form multiplets that
transform as irreducible representations of this unbroken $SO(4)$.
But $SO(4)$ multiplets also factorize into products of $SU(2)$
multiplets, and by acting with operator
$ K_{+} = \pi^{\dagger}_x - i \pi^{\dagger}_y  = -i
c_{\downarrow}^{\dagger} d_{\downarrow}^{\dagger} $ 
or
$ J_{+} = \pi_x - i \pi_y = -i
c_{\uparrow} d_{\uparrow}
$
one can move between states related by
holonomy. For example, states $\Phi_1^{\dagger} | \Omega \rangle $ and
$ \Phi_3^{\dagger} | \Omega \rangle $ form an $SO(4)$ multiplet that is a
doublet of the $J$ algebra and a singlet of the $K$ algebra, whereas the
states  $\Phi_2 | \Omega \rangle $ and
$ \Phi_3 | \Omega \rangle $ form an $SO(4)$ multiplet that is a
doublet of the $K$ algebra and a singlet of the $J$
algebra. Also since the two 
pairs belong to different multiplets holonomy does not mix them.

\subsection{SU(2) Holonomy of BCS and SDW Quasi-particles}
\label{BCS-SDW-holonomy}

Having illustrated the physical interpretation of the $SU(2)$ holonomy
in a $2$ site problem, we are now ready to investigate the physical
interpretation in terms of SDW and BCS quasi-particles. We start from
the tight-binding Hamiltonian with $SO(5)$ symmetric vector
interaction. We write it in spinor representation following notations
in \cite{prl98-1}
\begin{eqnarray}
{\cal H}(t) = &&{\sum_{{\bf p}\alpha}}'  \epsilon_{\bf p} ( \Psi_{{\bf
  p}\alpha}^{\dagger} \Psi_{{\bf p}\alpha} - 
  \Psi_{{\bf p}+{\bf Q}\alpha}^{\dagger} 
\Psi_{{\bf p}+{\bf Q}\alpha} )  \nonumber\\
&& \hspace{1cm} +~{\sum_{{\bf p}\alpha\beta}}' \Delta_{\bf
  p}~ n^a (t)~ ( \Psi_{{\bf 
  p}\alpha} \Gamma^a_{\alpha\beta} \Psi_{{\bf p}+{\bf Q}\beta} + 
\Psi_{{\bf p}+{\bf Q}\alpha}^{\dagger} \Gamma^a_{\alpha\beta}
  \Psi_{{\bf p}\beta} ) 
\label{meanfield}
\end{eqnarray}
Here 
\begin{eqnarray}
\Psi_{{\bf p}}^{T} = \{ c_{{\bf p}\uparrow}, c_{{\bf
p}\downarrow}, \phi_{\pi}({\bf p})~ c^{\dagger}_{-{\bf p}+{\bf
Q}\uparrow}, \phi_{\pi}({\bf p})~ c^{\dagger}_{-{\bf p}+{\bf 
Q}\downarrow} \} 
\end{eqnarray}
is the $SO(5)$ spinor with
$\phi_{\pi}({\bf p}) = sgn( cos p_x - cos p_y )$ and prime over
${\sum_{\bf p}}'$ denotes summation over half of the Brillouin
zone). When 
$n_a$ points in the $2,3,4$ directions, this is nothing but 
the SDW mean field Hamiltonian. When $n_a$ points in the $1,5$
direction, it represents the BCS mean field Hamiltonian. We are 
interested here in the case where the $n_a(t)$ field is time dependent.
This situation would arise naturally when one performs a 
Hubbard Stratonovich decomposition of the interaction term, and $n_a(t)$
is a dynamic mean field. As the $n_a(t)$ field fluctuates in time, the
quasi-particles ``rotate" between SDW and BCS characters. Here we 
would like to consider the adiabatic limit of this fluctuation and
identify the associated holonomy between SDW and BCS quasi-particles.

Following the discussions in sections \ref{SpinorRot}, we decompose
$\Psi_{{\bf p}\alpha}$ as 
\begin{eqnarray}
\Psi_{{\bf p}\alpha} (t) = S_{\alpha\beta}(t) \Phi_{{\bf p}\beta}(t)
\label{PsiPhieq}
\end{eqnarray}
where the spinor rotation matrix $S_{\alpha\beta}$ satisfies the same 
conditions (\ref{S-condition1}) and (\ref{S-condition2}) as before.
Substituting this decomposition into (\ref{meanfield}), we find the
Hamiltonian in the ``rotating frame":
\begin{eqnarray}
{\cal H}_0 = {\sum_{{\bf p}\alpha}}'  \epsilon_{\bf p} ( \Phi_{{\bf
    p}\alpha}^{\dagger} \Phi_{{\bf p}\alpha} - \Phi_{{\bf p}+{\bf
    Q}\alpha}^{\dagger}  
\Phi_{{\bf p}+{\bf Q}\alpha} ) + {\sum_{{\bf p}\alpha\beta}}' \Delta_{\bf
    p}~ ( \Phi_{{\bf p}\alpha}^{\dagger} 
    \Gamma^4_{\alpha\beta} \Phi_{{\bf p}+{\bf Q}\beta} + 
\Phi_{{\bf p}+{\bf Q}\alpha}^{\dagger} \Gamma^4_{\alpha\beta}
    \Phi_{{\bf p}\beta} ) 
\end{eqnarray}
This Hamiltonian can be easily diagonalized by a $SO(5)$ generalization
of the Bogoliubov transformation:
\begin{eqnarray}
\gamma^c_{{\bf p} \uparrow} &=  u_{\bf p} \Phi_{{\bf p}1} +  v_{\bf p}
\Phi_{{\bf p}+{\bf Q}1} \hspace{1cm}
\gamma^{v\dagger}_{-{\bf p} \uparrow} &=  u_{\bf p} \Phi_{{\bf p}3} +  v_{\bf p}
\Phi_{{\bf p}+{\bf Q}3}
\nonumber\\ 
\gamma^c_{{\bf p} \downarrow} &=  - u_{\bf p} \Phi_{{\bf p}2} +  v_{\bf p}
\Phi_{{\bf p}+{\bf Q}2} \hspace{1cm}
\gamma^{v\dagger}_{-{\bf p} \downarrow} &=  - u_{\bf p} \Phi_{{\bf
    p}4} +  v_{\bf p} \Phi_{{\bf p}+{\bf Q}4}
\nonumber\\ 
\gamma^v_{{\bf p} \uparrow} &=  v_{\bf p} \Phi_{{\bf p}1} - u_{\bf p}
\Phi_{{\bf p}+{\bf Q}1} \hspace{1cm}
\gamma^{c\dagger}_{-{\bf p} \uparrow} &=  v_{\bf p} \Phi_{{\bf p}3} -
u_{\bf p} \Phi_{{\bf p}+{\bf Q}3} 
\nonumber\\ 
\gamma^v_{{\bf p} \downarrow} &=  v_{\bf p} \Phi_{{\bf p}2} +  u_{\bf p}
\Phi_{{\bf p}+{\bf Q}2} \hspace{1cm}
\gamma^{c\dagger}_{-{\bf p} \downarrow} &=  v_{\bf p} \Phi_{{\bf p}4}
+  u_{\bf p} \Phi_{{\bf p}+{\bf Q}4}
\end{eqnarray}
where $v_{\bf p} = \sqrt{ \frac{1}{2} \left( 1 - \epsilon_{\bf p} / E_{\bf
      p} \right) }$,~ $u_{\bf p} = \sqrt{ \frac{1}{2} \left( 1 +
      \epsilon_{\bf p} / E_{\bf 
      p} \right) }$ and $E_{\bf p} = \sqrt{ \epsilon^2_{\bf p} +
      \Delta^2_{\bf p} }$~. We obtain
\begin{eqnarray}
{\cal H}_0 = \sum_{{\bf p}\sigma} E_{{\bf p}} \left(
\gamma^{c\dagger}_{{\bf p} \sigma} \gamma^{c}_{{\bf p} \sigma}
- \gamma^{v\dagger}_{{\bf p} \sigma} \gamma^{v}_{{\bf p} \sigma}
\right) 
\end{eqnarray}
and the ground state $|\Omega\rangle$ is obtained by filling the
negative energy states: 
\begin{eqnarray}
\gamma^c_{\pm {\bf p} \sigma} |\Omega\rangle = 0  \hspace{1.5cm}
\gamma^{v\dagger}_{\pm {\bf p} \sigma}  |\Omega\rangle = 0
\end{eqnarray}
Excited states
\begin{eqnarray}
\gamma^{c\dagger}_{\pm{\bf p} \sigma} |\Omega\rangle 
\hspace{1cm}
\gamma^v_{\pm{\bf p} \sigma} |\Omega\rangle 
\end{eqnarray}
are degenerate with energy $E_{\bf p}$. If at $t=0$ the
superspin $n_a$ 
points towards $n_4$, these quasiparticle states are readily expressed
in terms of 
electron creation/annihilation operators
\begin{eqnarray}
\gamma^{c\dagger}_{\pm {\bf p} \sigma} |\Omega\rangle = \left( \sigma~
u_{\bf p} c^{\dagger}_{\pm{\bf p} 
\sigma} + v_{\bf p} c^{\dagger}_{\pm{\bf p}+ {\bf Q} \sigma} \right)
|\Omega\rangle 
\nonumber\\  
\gamma^v_{\pm{\bf p} \sigma} |\Omega\rangle  =  \left( v_{\bf
    p} c_{\pm{\bf p} \sigma}   
- \sigma~ u_{\bf p} c_{\pm{\bf p} +{\bf Q} \sigma} \right)
|\Omega\rangle  
\end{eqnarray}
Elementary excitations in the original frame are
constructed as $\Phi_{{\bf p} \alpha} = S^{\dagger}_{\alpha\beta}(t)
\Psi_{{\bf p} \beta}$ where in the adiabatic approximation matrix
$S_{\alpha\beta}$ is taken from (\ref{Scomb}).

Now we imagine that $n_a(t)$ makes a closed adiabatic rotation
as discussed in section \ref{2sitesection}. It starts from $n_4$ and
returns at $t=T$ to the same direction. However, the matrix $S$ at
time $T$ 
does not have to return to unity and it may come back in the form
given in (\ref{S-evolution}). Equation (\ref{PsiPhieq})
then instantly tells us that after such cyclic rotation of the
superspin the quasiparticle state 
$\gamma^{c\dagger}_{{\bf p} \uparrow} 
|\Omega\rangle$ interchanges with $\gamma^v_{-{\bf p}  \uparrow }
|\Omega\rangle$, and  $\gamma^{c\dagger}_{{\bf p} \downarrow}
|\Omega\rangle$ interchanges with $\gamma^v_{-{\bf p}  \downarrow}
|\Omega\rangle$. This means that after superspin completes a revolution,
quasiparticle in the conductance band returns as a missing
quasiparticle in the valence band with the same spin but opposite
momentum. SDW quasiparticle and SDW quasihole have been interchanged!
What has been emitted in such process 
is an object with spin 1 and charge $2$
\begin{eqnarray}
\pi_{{\bf p}\sigma} = \gamma^{c}_{-{\bf p} \sigma}
\gamma^{v}_{{\bf p} \sigma} =  \phi_{\pi}({\bf p})~c_{-{\bf p}+{\bf Q}
  \sigma} c_{{\bf p} \sigma} +\dots
\end{eqnarray}
When summed over all ${\bf p}$'s the terms in $\dots$ disappear and we
get one of generators of $SO(5)$ algebra, the $\pi$
operator \cite{science,prl95}. So we obtained again  
that adiabatic transport of the superspin leads 
to a $SO(5)$ generalization of Andreev reflection, in which a $\pi$
particle has been emitted.

Interpretation of this result using unbroken $SO(4)$ symmetry of
Hamiltonian (\ref{meanfield}) is straightforward. When $n_a$ points
along $n_4$, the  
important generators of this unbroken symmetry ( see discussion after
equation (\ref{SO4})~) are
\begin{eqnarray}
K_{+} &=& \frac{1}{2} ( L_{21} - L_{35} + i L_{13} - i L_{25} )
\nonumber\\
&=& {\sum_{{\bf p}}} \phi_{\pi}({\bf p})
c_{{\bf p}\downarrow}^{\dagger} c_{-{\bf p}+{\bf
    Q}\downarrow}^{\dagger} =  {\sum_{{\bf p} }}'
\gamma^{v\dagger}_{{\bf p}\downarrow} \gamma^{c\dagger} _{-{\bf
    p}\downarrow} \nonumber\\
J_{+} &=& \frac{1}{2} ( L_{21} + L_{35} + i L_{13} + i L_{25} )
\nonumber\\
&=& {\sum_{\bf p}} \phi_{\pi}({\bf p}) c_{-{\bf p}+{\bf Q}\uparrow}
c_{-{\bf p}\uparrow}  =  {\sum_{{\bf p}}}'
\gamma^{c}_{-{\bf p}\uparrow}\gamma^{v}_{{\bf p}\uparrow} 
\end{eqnarray}
These are $\pi$ operators that move us between pairs of states that
form multiplets of the unbroken $SO(4)$.
\begin{eqnarray}
\gamma^{v}_{{\bf p}\downarrow} | \Omega \rangle
\stackrel{K_{+}}{\rightarrow}  \gamma^{c\dagger}_{-{\bf p}\downarrow} |
\Omega \rangle \nonumber\\
\gamma^{v}_{{\bf p}\uparrow} | \Omega \rangle
\stackrel{J_{-}}{\rightarrow}  \gamma^{c\dagger}_{-{\bf p}\uparrow} |
\Omega \rangle \nonumber\\
\gamma^{v}_{-{\bf p}\downarrow} | \Omega \rangle
\stackrel{K_{+}}{\rightarrow}  \gamma^{c\dagger}_{{\bf p}\downarrow} |
\Omega \rangle \nonumber\\
\gamma^{v}_{-{\bf p}\uparrow} | \Omega \rangle
\stackrel{J_{-}}{\rightarrow}  \gamma^{c\dagger}_{{\bf p}\uparrow} |
\Omega \rangle
\end{eqnarray}

Arguments presented in this Section clarify the origin
of the non-Abelian holonomy of spinors in the $SO(5)$ theory. 
It comes from a  non-zero overlap between SDW and BCS quasiparticle
states.

\section{Fermions and the SO(5) Non-Linear $\sigma$ Model }

We can now put our knowledge to a good use by considering how fermions
can be added to
the $SO(5)$ Non-Linear $\sigma$ Model. We start with a microscopic
Hamiltonian \cite{prl98-1} with $SO(5)$ symmetric pseudo-vector
interaction ( spinor notations of \cite{prl98-1} are used again ):
\begin{eqnarray}
{\cal H} &=& -t \sum_{x, \delta_{\mu}} \Psi^{\dagger}(x)
\Psi(x+\delta_{\mu}) + V \sum_x 
\hat{\Delta}^{\dagger}_a(x) \hat{\Delta}_a(x)\nonumber\\
\hat{\Delta}_a(x) &=& \sum_{\delta_{\mu}} (-)^{\mu}
\Psi(x) R \Gamma^a \Psi(x+\delta_{\mu})
\label{microscopicH}
\end{eqnarray}
where $\delta_{\mu}$ is a translation by one lattice constant in direction
$\mu$, and $(-)^{\mu}=1$ when $\mu=x$ and $-1$ when $\mu=y$. 
In momentum space such interaction corresponds to 
\begin{eqnarray}
{\cal H}_{int} &=& V \sum_{\bf q} \hat{\Delta}_a^{\dagger}({\bf q})
\hat{\Delta}_a ({\bf q}) 
\nonumber\\
\hat{\Delta}_a({\bf q}) &=& 2  \sum_{\bf k} w_{\bf k}
\Psi_{{\bf k}+{\bf q}} 
R \Gamma^a \Psi_{-{\bf k}} 
\end{eqnarray}
with $ w_{\bf k} = cos(k_x) - cos(k_y) $.
The mean field of this Hamiltonian in direction $\hat{\Delta}_1$ readily
reproduces the usual d-wave superconductivity. Antiferromagnetic
phase is less conventional since the  $SO(5)$
symmetry dictates it to have nodes in the energy gap, with no sign
change of the order parameter (see \cite{prl98-1} for a 
more detailed discussion).

We can use Hubbard-Stratonovich transformation to write Hamiltonian 
(\ref{microscopicH}) as  
\begin{eqnarray}
{\cal H} &=& -t \sum_{x,\delta_{\mu}} \Psi^{\dagger}(x)
\Psi(x+\delta_{\mu}) +  
\frac{1}{2V} \sum_x  \Delta^{\dagger}_a(x)  \Delta_a(x)   \nonumber \\
&& +  \sum_x
\Delta^{\dagger}_a (x) \sum_{\delta_{\mu}} (-)^{\mu}
\Psi(x) R \Gamma^a \Psi(x+\delta_{\mu}) + h.c.
\label{HSHamiltonian}
\end{eqnarray}
A simple Hartree-Fock saddle point for this Hamiltonian has a disadvantage
that it does not reproduce the non-linear $\sigma$ model directly. But we
can do better than the mean-field if we explicitly separate degrees of
freedom that contain Goldstone bosons of the $SO(5)$ rotations (see
\cite{Schulz1,Schulz2} for analysis of the $SO(3)$ symmetry in the
Hubbard model and \cite{nodons} for analysis of $U(1)$ symmetry and
nodal quasiparticles in $d$-wave superconductors ). It is important to
note that in the long-wavelength limit the  fields $\Delta_a(x)$
are real ( $\Delta^{\dagger}_a ( {\bf q}=0 ) = \Delta_a ( {\bf q}=0 )
$ ), so we can use matrix $\tilde{S}$ from Section \ref{SpinorRot} 
to rotate all spinors locally to point in one direction,  i.e. we
introduce new spinors 
$\tilde{\Psi}$ as 
\begin{eqnarray}
\Psi(x) = \tilde{S}(x) \tilde{\Psi}(x)
\end{eqnarray}
where rotation matrix $\tilde{S}(x)$ satisfies two conditions
\begin{eqnarray}
\Delta_a(x) \tilde{S}^{\dagger}(x) \Gamma^{a} \tilde{S}(x) &=&
\Gamma^4  \Delta(x)   \nonumber\\
\tilde{S}^{\dagger}(x) \tilde{S}(x) &=& 1 
\label{Ueqn}
\end{eqnarray}
with $\Delta^2(x) = \sum_a \Delta^{\dagger}_a(x) \Delta_a(x)  $.
One can
easily prove for $\tilde{S}$ the identity 
\begin{eqnarray}
\tilde{S}^{T} R \tilde{S} = R
\label{UTRUidntity}
\end{eqnarray} 
and use it to rewrite the first equation of (\ref{Ueqn}) as
\begin{eqnarray}
\Delta_a(x) \tilde{S}^{T}(x) R \Gamma^a \tilde{S}(x) = R \Gamma^4
\Delta(x)  
\label{UUTidentity}
\end{eqnarray}
Neglecting the high energy amplitude fluctuations of $\Delta(x)$ we can
express Hamiltonian (\ref{HSHamiltonian}) as  
\begin{eqnarray}
{\cal H} &=& {\cal H}_{MF} + {\cal H}_1 + {\cal H}_2
\nonumber\\
{\cal H}_{MF} &=& -t \sum_{x,\delta_{\mu}} \tilde{\Psi}^{\dagger}(x)
\tilde{\Psi}(x+\delta_{\mu}) +  \Delta   \sum_{x,\delta_{\mu}} (-)^{\mu}
\tilde{\Psi}(x) R \Gamma^4
\tilde{\Psi}(x+\delta_{\mu}) 
+ h.c. \nonumber \\
{\cal H}_1 &=& -t \sum_{x,\delta_{\mu}}  \tilde{\Psi}^{\dagger}(x)
\left( \tilde{S}^{\dagger}(x) \tilde{S}(x+\delta_{\mu}) -1 \right)
\tilde{\Psi}(x+\delta_{\mu}) + 
h.c. \nonumber \\
{\cal H}_2 &=&  \sum_{x,\delta_{\mu}} (-)^{\mu}
\tilde{\Psi}(x) \left( \tilde{S}^{T}(x) R \Gamma^a \tilde{S}(x+\delta_{\mu})
  \Delta_a(x) - R \Gamma^4  \Delta  \right) 
\tilde{\Psi}(x+\delta_{\mu}) + h.c.
\end{eqnarray}
It is helpful to define $\tilde{\Psi}(x) = G \tilde{\Phi}(x)$. Then
one can use 
\begin{eqnarray}
GR\Gamma^4G=\left( \begin{array}{cc} i \sigma_y & 0 \\ 0 & - i
    \sigma_y \end{array} \right)
\label{GRGidentity}
\end{eqnarray}
to write the mean-field Hamiltonian as
\begin{eqnarray}
{\cal H}_{MF} &=& {\cal H}_{+} + {\cal H}_{-} \nonumber\\
{\cal H}_{+} &=& \sum_{\bf k} \left\{ \epsilon_{\bf k} ( \tilde{\Phi}^{\dagger}_{{\bf k}1}
\tilde{\Phi}_{{\bf k}1} + \tilde{\Phi}^{\dagger}_{{\bf k}2}
\tilde{\Phi}_{{\bf k}2} ) 
+  \Delta w_{\bf k} ( \tilde{\Phi}_{k1}\tilde{\Phi}_{-{\bf k}2}
- \tilde{\Phi}_{{\bf k}2}\tilde{\Phi}_{-{\bf k}1} 
+  \tilde{\Phi}^{\dagger}_{{\bf k}2}  \tilde{\Phi}^{\dagger}_{-{\bf k}1}
-  \tilde{\Phi}^{\dagger}_{{\bf k}1}  \tilde{\Phi}^{\dagger}_{-{\bf k}2}
) \right\} \nonumber\\
{\cal H}_{-} &=& \sum_{\bf k} \left\{ \epsilon_{\bf k} ( \tilde{\Phi}^{\dagger}_{{\bf k}3}
\tilde{\Phi}_{{\bf k}3} + \tilde{\Phi}^{\dagger}_{{\bf k}4}
\tilde{\Phi}_{{\bf k}4} ) 
-  \Delta w_{\bf k} ( \tilde{\Phi}_{{\bf k}3}\tilde{\Phi}_{-{\bf k}4}
- \tilde{\Phi}_{{\bf k}4}\tilde{\Phi}_{-{\bf k}3} 
+  \tilde{\Phi}^{\dagger}_{{\bf k}4}  \tilde{\Phi}^{\dagger}_{-{\bf k}3}
-  \tilde{\Phi}^{\dagger}_{{\bf k}3}  \tilde{\Phi}^{\dagger}_{-{\bf k}4}
) \right\}
\label{HMeanField} 
\end{eqnarray}
with $\epsilon_{\bf k} = -2 t ( cos k_x + cos k_y ) $.
We introduce 
$\Phi_{+,{\bf k}}^{T} = \{ \tilde{\Phi}_{{\bf k}1},
\tilde{\Phi}_{{\bf k}2}, - \tilde{\Phi}^{\dagger}_{-{\bf k}2},
\tilde{\Phi}^{\dagger}_{-{\bf k}1} \}$ and 
$\Phi_{-,{\bf k}}^{T} = \{ \tilde{\Phi}_{{\bf k}3},
\tilde{\Phi}_{{\bf k}4}, - \tilde{\Phi}^{\dagger}_{-{\bf k}4},
\tilde{\Phi}^{\dagger}_{-{\bf k}3} \}$
and write equation (\ref{HMeanField}) as  
\begin{eqnarray}
{\cal H}_{\pm} &=& {\sum_{{\bf k}}}' \Phi_{\pm,{\bf k}}^{\dagger} (
\epsilon_{\bf k} \tau_3 \pm  2 \Delta w_{\bf k} 
\tau_1 ) \Phi_{\pm,{\bf k}}
\label{MF-short}
\end{eqnarray}
where we used $\tau$ matrices 
\begin{eqnarray}
\tau_1 = \left( \begin{array}{cc} 0 & \hat{1} \\ \hat{1} & 0
\end{array} \right)
\hspace{2cm}
\tau_3 = \left( \begin{array}{cc}\hat{1} & 0 \\ 0 & -\hat{1}
  \end{array} \right)
\end{eqnarray}
and later we will also use $\tilde{\sigma}_{\lambda}$ matrices
defined as 
\begin{eqnarray}
\tilde{\sigma}_{\lambda} = 
\left( \begin{array}{cc} \sigma_{\lambda} & 0 \\ 0 & \sigma_{\lambda}
 \end{array} \right) 
\end{eqnarray}

We expand (\ref{MF-short})  around ${\bf K}_1 =
(\frac{\pi}{2},\frac{\pi}{2})$  
and use $\epsilon_{{\bf K}_1+{\bf q}} \approx -2 t ( q_x + q_y )$,
$ w_{{\bf K}_1+{\bf q}} \approx q_y - 
q_x $ with analogous expressions for ${\bf K}_2 = (-
\frac{\pi}{2},\frac{\pi}{2})$. 
If we now define
$\Phi_{\pm, {\bf K}_i+{\bf q}} = \phi_{\pm i{\bf q}} $ we can write the
mean-field 
Hamiltonian as
\begin{eqnarray}
{\cal H}_{MF} &=& \sum_{{\bf q}, r = \pm } \phi_{r 1{\bf q}}^{\dagger}
  \left( 2t (q_x + q_y ) 
  \tau_3 + r \Delta ( - q_x + q_y) \tau_1 \right) \phi_{r 1{\bf q}}
  \nonumber \\ 
&+& \sum_{{\bf q}, r = \pm } \phi_{r2{\bf q}}^{\dagger} \left( 2t
  (-q_x + q_y ) 
  \tau_3 + r \Delta ( q_x + q_y ) \tau_1 \right) \phi_{r 2{\bf q}}  
\label{MFladder}
\end{eqnarray} 
It is convenient to introduce new coordinates
$ x_{1,2} = \frac{x \pm y }{\sqrt{2}} $ and velocities $ v_1 = 2
\sqrt{2} t$, $v_2 = 2 \sqrt{2} \Delta$, then from (\ref{MFladder}) we
obtain free Dirac Hamiltonian in $2+1$ dimension 
\begin{eqnarray}
{\cal H}_{MF} &=& \sum_{r=\pm} \int dx_1 dx_2~\{~ \phi^{\dagger}_{r1}
  \left( v_1 ( -i 
  \tau_3 ) \frac{\partial}{\partial x_1} + r v_2 ( -i
  \tau_1 ) \frac{\partial}{\partial x_2} \right)  \phi_{r1}
\nonumber \\  &+&
\phi^{\dagger}_{r2} \left( v_1 ( -i
  \tau_3 ) \frac{\partial}{\partial x_2} + r v_2 ( -i
  \tau_1 ) \frac{\partial}{\partial x_1} \right)  \phi_{r2}~ \}
\end{eqnarray}

Let us now look at ${\cal H}_1$. From equation (\ref{GStilda}) of
Section \ref{SpinorRot} 
we know that in the subspace of components 1 and 2 we have
$  \tilde{S}^{\dagger} \partial_{\mu} \tilde{S}~  = i
{\cal A}_{+,\mu}^{\lambda} \sigma_{\lambda} $, 
and in the subspace of components 3 and 4:~
$ \tilde{S}^{\dagger} \partial_{\mu} \tilde{S}~ = i
{\cal A}_{-,\mu}^{\lambda} \sigma_{\lambda}  
$. Here
${\cal A}_{\pm,\mu}^{\lambda} = \sum_{b=1}^{5} {\cal
  A}_{\pm}^{\lambda b}\partial_{\mu} n_{b}$ with ${\cal
  A}_{\pm}^{\lambda b}$ given by equation (\ref{calA+}) (see also  (\ref{imquat})).
Other elements of $  \tilde{S}^{\dagger} \partial_{\mu} \tilde{S} $
are zero in the adiabatic approximation.
Therefore,
${\cal H}_1$ can be expressed as
\begin{eqnarray}
{\cal H}_{1} &=& -t \sum_{x,\delta_{\mu}} \tilde{\Phi}^{\dagger}(x) G \left(
  \tilde{S}^{\dagger}(x) \tilde{S}(x+\delta_{\mu}) -1 \right) G
  \tilde{\Phi}(x+\delta_{\mu}) 
  \nonumber\\ 
 &=& -t \sum_{x,\delta_{\mu}} \tilde{\Phi}^{\dagger}(x) 
\left( \begin{array}{cc} i {\cal A}_{+,\mu}^{\lambda} \sigma_{\lambda} & 0 \\
0 &  i {\cal A}_{-,\mu}^{\lambda} \sigma_{\lambda} \end{array} \right) 
 \tilde{\Phi}(x+\delta_{\mu}) 
\label{H1}
\end{eqnarray}
 Taking the $\Phi_{+,{\bf k}}$
components of the last equation first, we have 
\begin{eqnarray}
\hspace{-1cm} {\cal H}_{1,+}\!&=&\!- 2 i t\!{\sum_{{\bf k} {\bf k}'
    {\bf q} \mu}}'\! 
\left( \tilde{\Phi}_{{\bf k}1}^{\dagger},
  \tilde{\Phi}_{{\bf k}2}^{\dagger}, - \tilde{\Phi}_{-{\bf k}2},
  \tilde{\Phi}_{-{\bf k}1}\right)\!\left( \begin{array}{cc} i {\cal
      A}_{+,\mu}^{\lambda}({\bf q}) \sigma_{\lambda} & 0 \\ 
        0  &  - i {\cal A}_{+,\mu}^{\lambda}({\bf q}) \sigma_{\lambda}
  \end{array} \right)\!\left( \begin{array}{c} \tilde{\Phi}_{{\bf
        k}'1} \\ \tilde{\Phi}_{{\bf k}'2} \\ 
 - \tilde{\Phi}^{\dagger}_{-{\bf k}'2} \\ \tilde{\Phi}^{\dagger}_{-{\bf k}'1}
\end{array} \right)\!sin ( k'_{\mu}) \delta({\bf k}-{\bf k}'+{\bf q})
 \nonumber\\  
&=&  2  t {\sum_{{\bf k} {\bf k}' {\bf q} \mu  }}'
\Phi_{+,{\bf k}}^{\dagger} 
   {\cal A}_{+,\mu}^{\lambda}({\bf q}) 
  \tilde{\sigma}_{\lambda}  \tau_3 \Phi_{+,{\bf k}'} sin(k'_{\mu})~ \delta({\bf
  k}-{\bf k}'+{\bf q}) 
\end{eqnarray}
Expanding the last expression around ${\bf K}_1$ and ${\bf K}_2$ we obtain 
\begin{eqnarray}
{\cal H}_{1+} &=&   2 t \sum_{{\bf q}_1 {\bf q}_2 {\bf q}}
\left({\cal A}_{+,x}^{\lambda}({\bf q}) + {\cal A}_{+,y}^{\lambda}({\bf q})\right) 
\Phi^{\dagger}_{+,{\bf K}_1+{\bf q}_1} \tilde{\sigma}_{\lambda} \tau_3
\Phi_{+,{\bf K}_1+{\bf q}_2} 
\delta({\bf q}_1-{\bf q}_2+{\bf q}) \nonumber \\ 
& +  & 2 t \sum_{{\bf q}_1 {\bf q}_2 {\bf q}}   \left( - {\cal
    A}_{+,x}^{\lambda}({\bf q}) + {\cal A}_{+,y}^{\lambda}({\bf q}) \right) 
\Phi^{\dagger}_{+,{\bf K_2}+{\bf q}_1} \tilde{\sigma}_{\lambda} \tau_3
\Phi_{+,{\bf K}_2+{\bf q}_2} 
\delta({\bf q}_1-{\bf q}_2+{\bf q}) \nonumber \\ 
&=& v_1 \int dx \left\{ \phi^{\dagger}_{+1}(x) ( -i
\tau_3 ) ( i  {\cal A}^{\lambda}_{+,1}(x) \tilde{\sigma}_{\lambda} ) \phi_{+1}(x)
+ \phi^{\dagger}_{+2}(x) ( -i
\tau_3 ) ( i  {\cal A}^{\lambda}_{+,2}(x) \tilde{\sigma}_{\lambda} ) \phi_{+2}(x)
\right\}
\label{H1gauge+}
\end{eqnarray}
where $ {\cal A}_{+,1}^{\lambda} = ( {\cal A}_{+,x}^{\lambda} + {\cal A}_{+,y}^{\lambda}
)/\sqrt{2} $ and 
$ A_{+,2}^{\lambda} = ( {\cal A}_{+,x}^{\lambda} - {\cal A}_{+,y}^{\lambda} )/\sqrt{2} $.
Analogously, for the $\Phi_{-,{\bf k}}$ components of ${\cal H}_1$ we have  
\begin{eqnarray}
{\cal H}_{1-}  &=& v_1 \int dx \left\{ \phi^{\dagger}_{-1}(x) ( -i
\tau_3 ) ( i  {\cal A}^{\lambda}_{-,1}(x) \tilde{\sigma}_{\lambda} ) \phi_{-1}(x)
+ \phi^{\dagger}_{-2}(x) ( -i
\tau_3 ) ( i  {\cal A}^{\lambda}_{-,2}(x) \tilde{\sigma}_{\lambda} ) \phi_{-2}(x)
\right\}
\label{H1gauge-}
\end{eqnarray}

Similar manipulations may be performed for ${\cal H}_2$. One uses
equations (\ref{Ueqn}), (\ref{UUTidentity}),  and
(\ref{GRGidentity}) to prove the identity 
\begin{eqnarray}
G \left(~ \tilde{S}^{T}(x) R \Gamma^a \tilde{S}(x+\delta_{\mu})~
  \Delta^a(x) - R \Gamma^4 \Delta 
~\right) G~ &=& \Delta~ G R \Gamma^4 G~ G~ (~ \tilde{S}^{\dagger}(x)
  \tilde{S}(x+\delta_{\mu}) -1~ )~ G\nonumber\\ 
&=& \Delta \left( \begin{array}{cc}  - {\cal A}^{\lambda}_{+,\mu} \sigma_y
  \sigma_{\lambda} & 0  \\
  0 & {\cal A}^{\lambda}_{-,\mu} \sigma_y \sigma_{\lambda}  \end{array} \right)
\end{eqnarray}
and then follows the steps that lead to 
(\ref{H1gauge+}) and (\ref{H1gauge-}). After straightforward
manipulations we obtain 
\begin{eqnarray}
{\cal H}_{2} &=& 2 \Delta {\sum_{{\bf k} {\bf k}' {\bf q} \mu}}'
\left\{ \Phi_{+,{\bf k}}^{\dagger} 
   {\cal A}_{+,\mu}^{\lambda}({\bf q}) 
  \tilde{\sigma}_{\lambda}  \tau_1 \Phi_{+,{\bf k}'} 
- \Phi_{-,{\bf k}}^{\dagger} 
   {\cal A}_{-,\mu}^{\lambda}({\bf q}) 
  \tilde{\sigma}_{\lambda}  \tau_1 \Phi_{-,{\bf k}'} 
\right\} sin(k'_{\mu})~ \delta({\bf
  k}-{\bf k}'+{\bf q}) 
\nonumber\\
 &=& v_2 \sum_{r=\pm} r \int dx \left\{ \phi^{\dagger}_{r1}(x) ( -i 
\tau_1 ) ( i  {\cal A}^{\lambda}_{r,2}(x) \tilde{\sigma}_{\lambda} ) \phi_{r1}(x)
+ \phi^{\dagger}_{r2}(x) ( -i
\tau_1 ) ( i  {\cal A}^{\lambda}_{r,1}(x) \tilde{\sigma}_{\lambda} ) \phi_{r2}(x)
\right\}
\label{H2gauge}
\end{eqnarray}

Combining all the pieces together we obtain
\begin{eqnarray}
{\cal H} = \sum_{r=\pm} \int dx_1 dx_2 \{~&v_1&~
\phi_{r1}^{\dagger}(x) ( -i \tau_3 ) [ 
\frac{\partial}{\partial x_1} +i {\cal A}_{r,1}^{\lambda} \tilde{\sigma}_{\lambda} ]
\phi_{r1} (x)  
\nonumber\\ +~ r &v_2&~ \phi_{r1}^{\dagger}(x) ( -i \tau_1 ) [
\frac{\partial}{\partial x_2} +i {\cal A}_{r,2}^{\lambda} \tilde{\sigma}_{\lambda} ]
\phi_{r1} (x)   
\nonumber\\ +~ &v_1&~ \phi_{r2}^{\dagger}(x) ( -i \tau_3 ) [
\frac{\partial}{\partial x_2} +i {\cal A}_{r,2}^{\lambda} \tilde{\sigma}_{\lambda} ]
\phi_{r2} (x)  
\nonumber\\ +~ r &v_2&~ \phi_{r2}^{\dagger}(x) ( -i \tau_1 ) [
\frac{\partial}{\partial x_1} +i {\cal A}_{r,1}^{\lambda} \tilde{\sigma}_{\lambda} ]
\phi_{r2} (x)  ~\}
\label{CDirac}
\end{eqnarray}
It is interesting to notice that we have a two-dimensional
representation of the 
Dirac matrices for our Dirac fermions. So each of the $\phi_{r1}$ and
$\phi_{r2}$ are chiral fermions, but parity is not broken since
parity transformation (  $t \rightarrow t$, $y \rightarrow y$, $x
\rightarrow -x$ \cite{Tsvelik,Semenoff} ) simply interchanges $\phi_{r1}$
and $\phi_{r2}$. 

Combining equation (\ref{CDirac}) with the $HP(1)$ representation of
the $SO(5)$ non-linear $\sigma$ model (\ref{partitionQNLSM}), we see
that the 
$SO(5)$ theory including both fermions and bosonic superspin variables
can be completely formulated as a $2+1$ dimensional relativistic
$SU(2)$ gauge field theory, with two bosonic ``Higgs fields'' ${\bf q}_1$
and ${\bf q}_2$ and four flavors of Dirac fermions $\phi_{ra}$
($r=\pm$ and $a=1,2$) \cite{Gaugefield}. The  
gauge field does not have a kinetic term ${\cal F}_{\mu\nu}^2$ in the
ordered phase. However such a term would be generated in the quantum
or thermal disordered phase \cite{Polyakov}. This complete formulation
of the $SO(5)$ theory is a central result of this work.

\section{Conclusions}

Our work showed that the ``square root" of the $SO(5)$ theory,
{\it i.e.} the spinor sector, has
a fascinatingly rich internal structure. From the point of view of
the $SO(5)$ theory, the fermionic quasi-particles of the high $T_c$
superconductors are still the ordinary SDW and BCS quasi-particles. 
The novel aspect of this system lies in the interplay between them.
Our paper is a tale about the intimate relationship between
these two quasi-particles.
The path leading towards their union travels through some of the
most beautiful areas of modern mathematics, with $SO(5)$ symmetry being a
unifying central theme. This mathematical structure enables us to
formulate the complete $SO(5)$ theory, including both fermionic and
superspin variables.

In the following, we shall discuss some
possible extensions of our work.

{\it $SO(5)$ symmetry breaking:} In order for these mathematical ideas
to apply to the real systems, the effects of explicit $SO(5)$ symmetry
breaking has to be carefully addressed. As noticed in reference 
\cite{ladder}, the degeneracy of the $SO(5)$ spinor multiplet is much
more robust than the vector multiplet, therefore, the results obtained
in the work could have more general applicability beyond exact $SO(5)$
symmetric models. In a model 
with $SO(5)$ symmetry breaking, the superspin is no longer confined to
move on $S^4$, but will trace out a general trajectory in the five
dimensional space. However, as long as it does not pass the degeneracy
point at the origin, the topological nature of the holonomy ensures that
the results would still apply in this case.

{\it Non-Abelian Bohm Aharonov effect:} In principle, one could
construct
AF and SC heterostructures and manipulate the direction of the superspin 
by magnetic fields and currents. Since the $SU(2)$ gauge field is
explicitly
determined by the superspin direction, one could construct regions with
finite $SU(2)$ flux, split the quasi-particle beams around it and
observe 
their interference pattern. The Abelian Bohm-Aharonov interference
produces modulation of the particle density, while the non-Abelian
Bohm-Aharonov interference produces modulation of the {\it mixing ratio}
of the particles belonging to the $SU(2)$ doublet.  

{\it Non-trivial fermion numbers of topological defects:} Recently, many 
authors investigated various types of topological defects
in the $SO(5)$ theory\cite{vortex,prl98-2,goldbart1,junction2}.
These
defects usually involve non-trivial spatial variations of the superspin
direction, their associated $SU(2)$ gauge field could give rise to
non-trivial
fermion numbers around the defects\cite{niemi_review}.

{\it $SO(5)$ Andreev reflection:}  In Section \ref{holonomysection} we have
seen that propagation of quasiparticles through regions with
nonuniform direction of the superspin may lead to $SO(5)$ 
Andreev reflection, when SDW quasiparticles in the conduction band turn into 
SDW quasiholes in the valence band or vice versa. This process
is similar to Andreev reflection at the superconductor/normal metal
interfaces, with an important difference that the emitted
particle is
not a Cooper pair, but a $\pi$ pair, a triplet two particle
excitation. Such processes could lead to novel phenomena in
superconducting/antiferromagnetic heterostructures: the
appearance of resonant tunneling as in \cite{prl97-1} or the
possibility of new bound states in heterostructures and around topological
defects of $SO(5)$. Our work gives a natural $SO(5)$
generalization of the Bogoliubov deGennes  
formalism to treat general bound states in all these cases, with
Sommerfeld quantization condition for the existence of the bound state
\cite{kashiwaya}
becoming a $2\times2$ matrix equation due to an $SU(2)$ holonomy of
the $SO(5)$ spinors.

{\it Single particle spectra in the pseudogap regime:} Within the
$SO(5)$
theory, the pseudogap regime is interpreted as the fluctuation regime
of the $SO(5)$ superspin vector. The quasi-particles in this regime have
fluctuating SDW/BCS characters, and can be naturally treated within the
finite temperature formalism of Dirac fermions coupled to fluctuating 
$SU(2)$ gauge fields. Connections to the photoemission experiments in
the pseudogap regime could be made.

{\it Relationship to other works:} The $SU(2)$ holonomy of $BCS$ and $SDW$
quasi-particles discussed in this paper bears some formal resemblance to 
the Affleck-Marston $SU(2)$ symmetry\cite{affleck} in the flux phase and the $SU(2)$
gauge theory recently formulated by Lee, Nagaosa, Ng and Wen\cite{patrick}.
However, these two approaches are obviously motivated by very different 
physical pictures and have entirely different physical meanings. Nevertheless,
it would be useful to explore their formal connections.

Recently, Balents, Fisher and Nayak\cite{nodons} studied the evolution of the
gapless fermion spectrum around the $d$ wave nodes from the SC state to the
insulating state. If the insulating state in question is a AF state, the formalism
developed in this work might be related to some of their considerations.

High $T_c$ superconductivity involve strong electron correlations which usually makes
perturbative calculations difficult to carry out. On the other hand, topological
effects are robust and have general validity even in the strong interaction
regime. Our work attempts to lay down a mathematical foundation for future studies of
topological effects within the $SO(5)$ theory. 
We hope that Nature made use of these elegant
mathematical concepts in the high $T_c$ superconductors.
  
We would like to thank Prof. A. Auerbach, E. Fradkin and A. Zee for
useful discussions. 
This work is supported by the NSF under grant numbers DMR-9400372 
and DMR-9522915.


\begin{thebibliography}{10}
\bibitem{Gaugefield}
Although $HP(1)$ representation in~eq. (58) gives explicitely only the
$A_{+,\mu}$ gauge 
  field, $A_{-,\mu}$ may be easily obtained following a procedure given~in
  Section~IIA.
\end{thebibliography}

\end{document}